\documentclass{aa}  

\usepackage{graphicx}
\usepackage{txfonts}
\usepackage{xcolor}
\usepackage{ulem}
\usepackage{pgffor}
\usepackage{xcolor}
\usepackage{rotating}
\usepackage{tikz}
\usepackage{morefloats}
\usepackage{morefloats}
\usepackage{graphicx}
\usepackage{caption}
\usepackage{subcaption}
\usepackage{csvsimple}
\usepackage{longtable}
\usepackage{booktabs}
\usepackage{pdfpages}
\usepackage{pgffor}
\usepackage{multicol}
\usepackage{footnote}
\usepackage{etoolbox}
\usepackage{array} 

\usepackage[pdftex]{hyperref}

\begin{document}

   \title{Investigating chemical variations between interstellar gas clouds in the Solar neighbourhood}

   \subtitle{}

   \author{T. Ramburuth-Hurt \inst{1}
   \and
           A. De Cia \inst{1,2}
    \and
            J.-K. Krogager \inst{3,4}
    \and
            C. Ledoux \inst{5}
    \and
            E. Jenkins \inst{6}
    \and
        A. J. Fox \inst{7,8}
           \and
        C. Konstantopoulou \inst{1}
    \and 
        A. Velichko \inst{1, 9}
    \and
        L. Dalla Pola \inst{1}
          }

   \institute{Department of Astronomy, University of Geneva, Chemin Pegasi 51, Versoix, Switzerland %1
    \and
        European Southern Observatory, Karl-Schwarzschild-Str. 2, 85748 Garching, Germany % 2
    \and
        Université Claude Bernard Lyon 1, Centre de Recherche Astrophysique de Lyon UMR5574, 9 Av. Charles André, 69230 Saint-Genis-Laval, France %3
    \and
        French-Chilean Laboratory for Astronomy (FCLA), CNRS-IRL3386, U. de Chile, Camino el Observatorio 1515, Casilla 36-D, Santiago, Chile % 4
    \and 
        European Southern Observatory, Alonso de Córdova 3107, Vitacura, Casilla 19001, Santiago, Chile %5
    \and
        Department of Astrophysical Sciences, Princeton University, Princeton, NJ 08544-1001, USA %7
    \and 
        AURA for ESA, Space Telescope Science Institute, 3700 San Martin Drive, Baltimore, MD 21218, USA %8
    \and
        Department of Physics \& Astronomy, Johns Hopkins University, 3400 N. Charles Street, Baltimore, MD 21218, USA %9
    \and 
        Institute of Astronomy, Kharkiv National University, 4 Svobody Sq., Kharkiv, 61022, Ukraine %10
             }

   \date{Received xx; accepted yy}

  \abstract
{The interstellar medium (ISM) is a fundamental component of the Milky Way. Studying its chemical composition and the level of its chemical diversity gives us insight into the evolution of the Milky Way and the role of gas in the Galactic environment. In this paper, we use a novel simulation technique to model the distribution of total hydrogen between gas components, and therefore derive new constraints on the dust depletion and metallicity. We study individual gas components along the lines of sight towards eight bright O/B stars within 1.1 kpc of the Sun using high-resolution HST/STIS absorption spectra (R $\sim$ 114 000). We measure the level of dust depletion for these individual components and find components with higher levels of dust depletion compared to Milky Way sightlines in the literature. We find large ranges in the level of dust depletion among components along lines of sight, up to a factor of 15 (or 1.19 dex). Although it is not possible to directly measure the metallicity of individual components due to the saturated and damped Lyman-$\alpha$ line, we investigate possible metallicity ranges for individual gas components by exploring many different distributions of the total hydrogen gas between components. We select possible combinations of these gas fractions which produce the minimum metallicity difference between components, and for these cases we determine individual metallicities to accuracies that range between $\sim$ 0.1 to 0.4 dex. This work shows that full line-of-sight analyses wash out the level of diversity along lines of sight, and that component-by-component studies give a more in-depth understanding of the chemical intricacies of the interstellar medium. }


   \maketitle
%
%-------------------------------------------------------------------
\section{Introduction}

    Studies of the chemical composition of the Milky Way's interstellar medium (ISM) give us important insight into its evolution and the role of gas in the Galactic environment. Determining the level of diversity in the metallicity of the warm neutral gas in the Solar neighbourhood is important for finding evidence of low-metallicity gas in-fall onto the Milky Way, for example from high-velocity clouds \citep[HVCs,][]{Richter2017, Fox+2019_HVCs}. Low-metallicity gas inflows from the circumgalactic medium (CGM) have been predicted in simulations \citep{Schaye+2015_EAGLE, Crain+2015_EAGLE, Nelson+2019_TNG50, Pillepich+2019_TNG50, Peroux+2020_met-ang-dep}. Evidence for pockets of low-metallicity gas in the ISM of the Milky Way could be an indication of low-metallicity gas in-fall \citep{DeCia+2021}.
    
    Absorption-line spectroscopy is a powerful method for studying the metal content of the gas in galaxies because it gives access to the column densities of many metals (ions), and the methods of measuring the column densities are very robust. Additionally, UV absorption-line spectroscopy is particularly valuable because the resonant lines of the most dominant ionisation states of the most abundant cosmic metals (O, C, Fe, Si, etc.) are available in the UV. There are several phenomena to take into account when calculating the total overall metallicity (i.e. gas + dust), including that of dust depletion, ionisation effects and nucleosynthesis effects. 

    Dust depletion is the phenomenon whereby metals are incorporated into dust grains and are no longer observable in the gas phase \citep{Field1974, Phillips+1982, Jenkins+1986, SavageSembach1996, Jenkins2009, DeCia+2016, DeCia+2021, Roman-Duval+2021, Konstantopoulou+2022}. The depletion of different metals correlates with each other to varying degrees depending on how easily they form dust. The tendency of a metal to deplete into dust is called its refractory index and it is broadly correlated to the condensation temperature of the metal \citep{SavageSembach1996, DeCia+2016, Konstantopoulou+2022}. Metals that form dust more readily are referred to as refractory, while those that form dust less easily are volatile. The measure of the relative abundances between metals with different refractory properties is therefore a measurement of the amount of dust depletion. In this method for characterising dust depletion, called the 'relative method' \citep{DeCia+2016, DeCia+2021, Konstantopoulou+2022}, the determination of the refractory indices of metals is based on the relative abundances of Zn and Fe. An advantage of the relative method is that it does not make any assumption on the metallicity of the gas.
    
    Other than dust depletion, there are at least two other phenomena that can affect the abundances of metals: ionisation and nucleosynthesis effects. In general, the ions that live in the warm neutral ISM of the Milky Way are singly ionised (with the exception of \ion{O}{I}), and are not highly susceptible to further ionisation because they are protected from ionising radiation by neutral hydrogen \ion{H}{I}. Elements that are further ionised from ionisation potentials higher than that of \ion{H}{I} (13.6eV) can also survive in \ion{H}{II} regions, which could result in their apparent over-abundance. For example, as shown in \cite{Jenkins2009}, the second ionisation potential of S is higher than that of most other elements, meaning that it can be present in a broader range of \ion{H}{II} regions (i.e. where the radiation field is harder and many other elements will be doubly ionised), so it is more likely for it to have contributions from \ion{H}{II} regions than most other singly ionised species. Nucleosynthesis effects, for example $\alpha$-element enhancement and Mn under-abundance, may also affect observed metal abundances. Although $\alpha$-element enhancements are observed in the Magellanic Clouds \citep{DeCia+2024} and in distant gas-rich galaxies \citep{Velichko+2024}, they are not generally seen in the Milky Way.
    
    In most cases, chemical enrichment and metallicities in the neutral ISM are determined for the whole line of sight towards stellar targets in the Milky Way \citep[e.g.][]{Jenkins2009, DeCia+2021, Ritchey+2023}. It is possible that intricacies at smaller scales, i.e. at the level of individual gas components along the lines of sight, make the ISM more complex at this level, and that studying whole lines of sight overlooks small-scale chemical variations. Evidence for this comes from deviations from the normal abundance patterns (the distribution of the observed [$X$/H] = $\log{\frac{N(X)}{N(H)}} - \log{\frac{N(X_{\odot})}{N(H_{\odot})}}$ against their refractory index for different metals). In these cases, volatile elements seem to deviate from the more typical linear shape of abundance patterns. A possible explanation for these abnormal abundance patterns is that the gas along the lines of sight is a mix of low- and high-depletion and/or metallicity.  
    
    With high-resolution absorption-line spectroscopy it is possible to dissect lines of sight, and identify and study the chemical properties of individual gas components \citep{Welty+2020_HD62542, Ramburuth-Hurt+2023}. This allows for a more comprehensive analysis of the variation (if any) in chemical enrichment between different gas clouds along lines of sight. A limitation of the component-by-component analysis is that it is not possible to know the H column density for individual components due to the Lyman-$\alpha$ line being damped, and therefore too broad for the separation of individual components. This means it is not possible to directly measure the metallicity of individual components [M/H]$_i$. It is still possible to characterise the chemical enrichment of individual components using dust depletion as a proxy \citep[e.g.][]{Welty&Crowther2010, Ramburuth-Hurt+2023}. However, although the level of dust depletion has some dependence on metallicity, it also depends on the physical conditions like temperature and density. In high pressure, high density and low temperature gas, such as in the Milky Way disk, the production of molecules and dust has a smaller dependence on metallicity \citep{Blitz&Rosolowsky2006_dust-metallicity, Balashev+2017_dust-metallicity}. This means that the distribution of dust depletion does not fully trace that of the metallicity \citep{Draine2003_ISMdust, Mattsson+2014_dustgrowth, Blitz&Rosolowsky2006_dust-metallicity, Balashev+2017_dust-metallicity}. 

    In this work, we study the dust depletion in individual gas components along eight lines of sight in the Solar neighbourhood with high-resolution UV spectra from HST/STIS. We present a novel methodology of investigating and constraining the metallicities for these individual components using simulations based on hydrogen gas fraction combinations. This paper is organised as follows: Section \ref{sec:data} introduces the data employed in this investigation, Section \ref{sec:methods} elaborates on our methodology, Section \ref{sec:results} presents and discusses the findings of this study, and Section \ref{sec:conclusions} summarises the conclusions drawn from our work.
 
%======================================================================
%======================================================================
%======================================================================

\section{Data} \label{sec:data}
In this work, we study the ISM along the lines of sight towards eight bright O/B-type stars within 1.1 kpc of the Sun. These targets are a sub-sample of the 25 targets compiled by \citet{DeCia+2021}. We select the sub-sample based on three main criteria:
   \begin{enumerate}
       \item The presence of multiple components in the \ion{Ti}{II} absorption lines, based on spectra from VLT/UVES and Kitt Peak. 
       \item An upturn of volatile elements in the abundance patterns in \citet{DeCia+2021} which suggests a mix of gases of different dust depletion and/or metallicities along the line of sight.
       \item Lower metallicity \citep[$\leq-0.16$ dex according to][]{DeCia+2021} over the integrated line of sight. 
   \end{enumerate} 

   While it is possible that the metallicities from \cite{DeCia+2021} are underestimated with respect to other strategies \citep{Ritchey+2023}, we based our sample selection on lines of sight that potentially showed signatures of low metallicity gas, which is the motivation for the third condition. The star type, distance from the Galactic center calculated from Gaia data release 3 (GR3), galactic latitude (b) and longitude (l), and the extinction ($A_\mathrm{V}$) for the eight targets are included in Table \ref{table:target-info}.

\begin{table*}
\centering
\caption{Target information with distances from the Galactic center calculated using Gaia DR3. The $A_V$ values are taken from \cite{Valencic+2004_RV}.}
\label{table:target-info}
\begin{tabular}{c|c|c|c|c|c|c|c}
\hline \hline
Target & HD & Type & Galactocentric & Heliocentric & $l$ ($^{\circ}$) & $b$ ($^{\circ}$) & $A_V$ (mag)  \\
&  & & distance (kpc) & distance (pc) &   & &   \\
\hline 
$\theta^1$ Ori C & 37022  & O6-7p           & 8.6 $\pm$ 0.2 & 399  $\pm$ 21  & 209.0  & $-$19.4 & 1.78 $\pm$ 0.36 \\
HD~110432        & 110432 & B0.5IVpe        & 8.1 $\pm$ 0.2 & 438  $\pm$ 14  & 302.0  & $-$0.2  & --  \\
$\rho$ Oph A     & 147933 & B2IV            & 8.2 $\pm$ 0.2 & 137  $\pm$ 2   & 353.7  & 17.7    & 2.58 $\pm$ 0.34  \\
$\chi$ Oph       & 148184 & B2IVpe          & 8.1 $\pm$ 0.2 & 152  $\pm$ 4   & 357.9  & 20.7    & -- \\
HD~154368        & 154368 & O9Ia            & 7.2 $\pm$ 0.2 & 1064 $\pm$ 38  & 350.0  & 3.2     & 2.53 $\pm$ 0.20 \\
$\kappa$ Aql     & 184915 & B0.5IIIn        & 7.9 $\pm$ 0.2 & 506  $\pm$ 39  & 31.8   & $-$13.3 & -- \\
HD~206267        & 206267 & O6.0V((f))+O9:V & 8.4 $\pm$ 0.2 & 735  $\pm$ 118 & 99.2   & 3.7     & 1.47 $\pm$ 0.14  \\
HD~207198        & 207198 & O9IIe           & 8.6 $\pm$ 0.2 & 1002 $\pm$ 34  & 103.1  & 7.0     & 1.50 $\pm$ 0.29 \\
\hline
\end{tabular}
\end{table*}

% -------------------------------------------------------------------------

\subsection{HST data}

The data we use in this work are primarily from the high spectral resolution ($R \sim 114~000$) HST/STIS program ID:16750 of the eight bright O/B stars within 1.1 kpc of the Sun. This program had 15 orbits in two settings: E140H with a central wavelength of 1271 $\AA$, covering 1170 -- 1372 $\AA$, and E230H with a central wavelength of 2113 $\AA$ covering 1974 -- 2251 $\AA$. In the case of $\rho$ Oph A, HD~206267 and HD~207198, the spectra for the E230H/2113 setting already exists in the archive. Table \ref{tbl:observation-details} contains observational details of this program including the number of orbits, exposure time and signal-to-noise ratio (SNR) for each setting. The wavelength bands were chosen to include the absorption lines for a number of transitions of interest tabulated in Table \ref{tbl:ion-transitions}. We use the 1D spectra directly from the archive, which have been reduced by the standard CALSTIS pipeline.

% -------------------------------------------------------------------------

\subsection{Auxiliary data}
We made use of the following data to supplement the absorption lines listed above. Specifically, because the decomposition of the lines is important for this work, the supplementary very high-resolution data are used to determine the substructure of the absorption lines.

\begin{itemize}
    \item Archival high-resolution HST/STIS spectra (R $\sim 114~000$, FWHM $\sim 2.63$ km~s$^{-1}$) that cover the \ion{Fe}{II} $\lambda\lambda$ 2260, 2367 lines for the targets $\rho$ Oph A, HD~206267 (ID:16285) and HD~207198 (ID:9465).
    
    \item VLT/ESPRESSO spectra (ID:0102.C-0699(A), $R \sim 190~000$, FWHM $\sim 1.54$ km~s$^{-1}$) that cover \ion{Ca}{II} $\lambda\lambda$ 3934, 3969, \ion{K}{I} $\lambda\lambda$ 7664, 7698, and \ion{Na}{I} $\lambda\lambda$ 5889, 5895.
    
    \item Kitt Peak National Observatory (KPNO) spectra from the Coude Feed Telescope with Camera 6 ($R \sim 220~000$, FWHM $\sim$ 1.35 km~s$^{-1}$) that cover \ion{Ca}{II} $\lambda$ 3934 and \ion{K}{I} $\lambda\lambda$ 7664, 7698, and from the Coude Feed Telescope with Camera 5 ($R \sim 88~000$, FWHM $\sim$ 3.4 km~s$^{-1}$) that covers \ion{Ti}{II} $\lambda$ 3384. 
    
    \item McDonald spectra ($R \sim 176~000$, FWHM $\sim 1.7$ km~s$^{-1}$) that cover \ion{K}{I} $\lambda\lambda$ 7664, 7698 and \ion{Ca}{II} $\lambda\lambda$ 3934, 3969 \citep{Pan+2004}.
    
    \item UHRF spectra covering \ion{Na}{I} $\lambda\lambda$ 5889, 5895, \ion{K}{I} $\lambda$ 7698 \citep{WeltyHobbs2001} and \ion{Ca}{II} $\lambda\lambda$ 3934, 3969 ($R \sim 880~000$, FWHM $\sim 0.34$) \citep{Price+2001}.
    
    \item VLT/UVES spectra (ID:194.C-0833, $R \sim 92~600$, FWHM $\sim 3.2$ km~s$^{-1}$) that cover \ion{Ti}{II} $\lambda\lambda\lambda$ 3230, 3242, 3384.
\end{itemize}

Table \ref{tbl:ion-transitions} contains the rest wavelengths, oscillator strengths and $\gamma$ values for each of the transitions we consider here. 

\begin{table}[]
\centering
\begin{tabular}{c|c|c|c}
\hline \hline
Ion & Transition & Oscillator & Reference \\ 
             & wavelength (\AA) & strength &  \\ \hline
\ion{Ni}{II} & 1317.217                    & 0.0596                &  [1]  \\
\ion{Cr}{II} & 2056.2568                   & 0.11                  &  [2] \\
\ion{Cr}{II} & 2062.2343                   & 0.078                 &  [3] \\
\ion{Cr}{II} & 2066.161                    & 0.0515                &  [3] \\
\ion{Fe}{II} & 2249.8754                   & 0.00182               &  [3] \\
\ion{Fe}{II} & 2260.7791                   & 0.00244               &  [3] \\
\ion{Mn}{II} & 1197.1843                   & 0.205                 &  [3] \\
\ion{Mn}{II} & 1199.3912                   & 0.16                  &  [3] \\
\ion{Mn}{II} & 1201.1177                   & 0.115                 &  [3] \\
\ion{Ge{II}} & 1237.0591                   & 1.230                 &  [3] \\
\ion{Si}{II} & 1260.4221                   & 1.45                  &  [3] \\
\ion{Si}{II} & 1304.3702                   & 0.106                 &  [3] \\
\ion{Mg}{II} & 1239.9253                   & 0.000633              &  [3] \\
\ion{Mg}{II} & 1240.3947                   & 0.000355              &  [3] \\
\ion{S}{II}  & 1250.578                    & 0.0052                &  [3] \\
\ion{S}{II}  & 1253.805                    & 0.01                  &  [3] \\
\ion{S}{II}  & 1259.518                    & 0.016                 &  [3] \\
\ion{Zn}{II} & 2026.137                    & 0.63                  &  [2] \\
\ion{Zn}{II} & 2062.6604                   & 0.31                  &  [2] \\
\ion{P}{II}  & 1301.8743                   & 0.0196                &  [4]\\
\ion{O}{I}   & 1302.168                    & 0.052                 &  [3] \\
\ion{O}{I}   & 1355.5977                   & 0.00000116            &  [3] \\
\ion{Ca}{II} & 3934.777                    & 0.65                  &  [3] \\
\ion{Ca}{II} & 3969.591                    & 0.32                  &  [3] \\
\ion{Ti}{II} & 3230.1219                   & 0.0773                &  [3] \\
\ion{Ti}{II} & 3242.9181                   & 0.233                 &  [3] \\
\ion{Ti}{II} & 3384.73                     & 0.353                 &  [3] \\ \hline
\end{tabular}
\caption{Rest wavelengths, oscillator strengths and their reference for the transitions considered in this work. [1] \cite{BoisseBergeron2019}, [2] \cite{Cashman+2017}, [3] \cite{Morton2003}, [4] \cite{Brown+2018}}
\label{tbl:ion-transitions}
\end{table}

%======================================================================
%======================================================================
%======================================================================

\section{Methodology} \label{sec:methods}
   
    To analyse the chemical enrichment and variations along these lines of sight, we start with measuring column densities using Voigt-profile fitting software packages, described in Section \ref{sec: measuring column densities}. In this work, we go beyond the more typical study of the abundance patterns (variations of [X/H] with the refractory index) along the full line of sight by performing a component-by-component analysis along the lines of sight. To do this, we construct metal patterns \citep[also called depletion patterns in ][]{Ramburuth-Hurt+2023}. These are the metal-to-metal variations with respect to their tendency to deplete onto dust grains, namely the refractory indices, plotted against their equivalent metal column densities ($\log{N(X)} - \log{N(H)}$ for a given metal $X$). From these metal patterns we can pinpoint the amount of dust depletion in individual components (Section \ref{sec:measuring dust depletion}). We then perform simulations, based on the observations, to further constrain the possible ranges of metallicities of the individual components (Section \ref{sec:simulating clouds}). We do this by exploring many distributions of the total hydrogen gas among the components. In this section, we describe this methodology in more detail and discuss some of its caveats.

% -------------------------------------------------------------------------
    \subsection{Measuring column densities} \label{sec: measuring column densities}
    
    The \ion{H}{I} column densities are measured from the Lyman-$\alpha$ absorption line at 1215 \AA\ by determining the radial velocity for most of the gas from the \ion{O{I} $\lambda1355$ line. This} is a good indicator of the velocity structure of \ion{H}{I} because its ionisation energy is very similar to that of \ion{H}{I} and the ionisation fractions of \ion{O}{I} and \ion{H}{I} are strongly coupled to each other by a charge exchange process with a large rate coefficient \citep{Stancil+1999, Spirko+2004}. This line is also far from being saturated, and oxygen is not much depleted onto dust grains. The outcome of this velocity measurement is used to establish an origin for trial \ion{H{I} absorption profiles.} The column density is then measured by determining the best fit damping profile for the Lyman-$\alpha$ line. In some cases, one side of the \ion{H}{I} absorption profile showed sufficient cancellation of the \ion{H}{I} absorption while the other side exhibited extra absorption. This effect could be due to some absorption by the stellar Lyman-$\alpha$ if the star's velocity is displaced from that of the foreground gas. It is not easy to confirm this interpretation, however, because there is a high probability that the star could be a spectroscopic binary and therefore could exhibit this same effect. The targets with this effect are $\theta^1$ Ori C, HD~154368, HD~110432, $\chi$ Oph. Column densities of H$_2$ are taken from \cite{Savage+1977}.
    
    We use two software packages for measuring the metal column densities for the targets in this work. For most of the targets ($\rho$ Oph A, HD~110432, HD~154368, HD~2062676 and HD~207198) we use the Voigt-profile fitting software \texttt{VoigtFit} \citep{Krogager2018}. For $\kappa$ Aql, $\chi$-Oph, $\theta^1$ Ori C, metal column densities are measured using \texttt{fits6p} \citep{Welty+1991, Welty+2003, Welty+2020_HD62542}. Note that the uncertainties on the column densities for the fits with \texttt{fits6p are underestimated.}
    
    For all targets, the structure of the absorption lines is determined by first fitting the lines covered by the highest-resolution spectra we have. We do this in order to best capture the intricacies of the line profiles, and because the STIS data are under-sampled (less than 2 pixels per resolution element). We assume that all species have a common velocity structure, which is obtained from the highest resolution spectra, but with possible slight adjustments based on the UV profiles, especially in the cases of weak components or broad lines. We assume that the component absorption lines are represented by symmetric Voigt profiles, and that asymmetry implies the presence of more than one component. 
    
    For $\rho$-Oph A, HD~110432, HD~154368, HD~2062676 we use one (or both simultaneously, when available) of \ion{Ca}{II} $\lambda$$\lambda$3934, 3969. For HD~207198, the \ion{Ca}{II} lines are saturated, and therefore cannot be used. Instead, we fit simultaneously to \ion{Cr}{II} $\lambda$$\lambda$$\lambda$ 2056, 2062, 2066 and \ion{Ti}{II} $\lambda$ 3384 to determine the structure. For $\kappa$ Aql, $\chi$ Oph and $\theta^1$ Ori C, we obtained a mutually consistent structure from the highest resolution \ion{K{I}, \ion{Na}{I}, and \ion{Ca}{II} spectra that were available.} The $b$-values (which represent the distribution of velocities due to gas turbulence and thermal kinematics, with turbulence dominating in the warm neutral ISM), velocities and column densities for fits to the highest resolution data are included in Table \href{https://zenodo.org/records/14793736}{C.1} in Appendix \href{https://zenodo.org/records/14793736}{C}.
    
    Once the component profile is determined, the decomposition is then assumed to be applicable to all other singly ionised species. The column densities are measured using all available (non-saturated, non-contaminated) lines from the same species according to this structure. In several cases, there is a velocity offset between the optical and UV spectra, which is taken into account when applying the velocity structure across these spectra. The spectra and fits are shown in Appendix \href{https://zenodo.org/records/14793736}{C}.

    For some of the ions, it is not possible to decompose their absorption lines to the same level of intricacy as the high resolution lines either because the lines are weaker, because of blending at lower resolution, or for both of these reasons. As a result, and to ensure the reliability of our measurements, we consider the column densities of groups of components (grouped by velocity) rather than all components individually, also because these are generally more well-constrained than individual components. It is important to note that the decomposition of components is not necessarily physical, and we do not know if the separation in velocity space means that they are separated in physical space. In this work, we are careful to group the components using the highest-resolution spectra to identify individual components. Tables C.2 -- C.9 in Appendix \href{https://zenodo.org/records/14793736}{C} present the column densities for groups of components, including the way that the components were grouped and their respective velocity ranges. For simplicity and to avoid repetitive phrasing, we use the term ``components" to refer to groups of gas components for the remainder of this paper.
    
    \ion{S}{II}, \ion{Si}{II} and \ion{Zn}{II} are often saturated. We do not use the \ion{S}{II} and \ion{Si}{II} lines in our analysis because of their severe saturation, and do not use \ion{Zn}{II} for HD~206267 and HD~207198. Some of the components on the edges of the \ion{Zn}{II} absorption lines are salvageable, and we take the column densities from \texttt{VoigtFit} to be lower limits on the total line of sight. When there are individual components on the edges of the \ion{Zn}{II} profiles that are unsaturated, we include these in the component-by-component analysis. 

    Most of the absorption lines towards $\rho$ Oph A are saturated (see the fits in Fig. \href{https://zenodo.org/records/14793736}{G.3}). We can recover reliable column densities for the first component only, but the lines for the second group of components (group 2, components 4 -- 7) are too saturated, with the exception of \ion{Ti}{II} and \ion{O}{I}. Therefore, for this target, we only take into account \ion{Ti}{II} and \ion{O}{I} for the metal pattern of group 2. We do not perform further analysis on this line of sight.
   
% -------------------------------------------------------------------------

   \subsection{Dust depletion of individual components} \label{sec:measuring dust depletion}
   
    We use the relative method of characterising dust depletion where the depletion of element $X$, denoted as $\delta_X$, is estimated from relative abundances \citep[][]{DeCia+2016, Konstantopoulou+2022}, not observed abundances. Relative abundances are defined as [X/Y] = $\log{N(X)/N(Y)} - \log{N(X)_{\odot}/N(Y)_{\odot}}$, where the Solar abundances are taken from \citep{Asplund+2021} following the recommendation of \citep{Lodders2009}. 
    
    The relative method is based on full line-of-sight measurements along lines of sight in the Milky Way, Magellanic Clouds, GRB- and QSO-DLAs. Here we assume that the method is applicable to the individual components that we study in this paper based on the following. Firstly, \cite{Ramburuth-Hurt+2023} demonstrated that the dust depletion sequences hold for individual components in QSO-DLAs, i.e. ionisation effects are limited, and the dominant effect of dust depletion can be clearly identified and studied also in individual components. However, that study did not include Milky Way lines of sight, or systems with a high level of dust depletion. This adds some level of uncertainty to this study. The line of sight towards HD~62542 in the Milky Way \citep{Welty+2020_HD62542} is an example of a case where the depletion patterns in individual components may depart from the general trends. Evidence for deviations from the linear behaviour of metal patterns would provide clues that this method is not applicable to individual components in this regime. Secondly, dust depletion has been shown to behave similarly in different environments \citep[QSO-, GRB-DLAs, Magellanic Clouds and Milky Way,][]{Konstantopoulou+2022}. Finally, the dust tracer [Zn/Fe] correlates with other relative abundances that are used as dust tracers, such as [Si/Ti] \citep{DeCia+2016, Konstantopoulou+2024_DUNE}. 
    
    Most extragalactic work is based on the full line-of-sight analysis of mid/high-resolution (R < 50 000) spectra. 
    
    The amount of depletion is expressed as
    \begin{equation} \label{eq:meth1}
        \delta_X = A2_X + B2_X \times \mathrm{[Zn/Fe]}_{\rm fit},
    \end{equation}
   where [Zn/Fe]$_{\rm fit}$ is a measure of the overall strength of depletion, or the depletion factor; $A2_X$ and $B2_X$ are coefficients of the depletion sequence fits in \citet{Konstantopoulou+2022} and are specific for each element (see Table \ref{table:depletion sequences}).  $B2_X$ is the refractory index, which represents the ease with which a metal is captured onto dust grains. Refractory elements form dust relatively readily and have more negative $B2_X$ values, while volatile elements do do not form dust so easily and have less negative $B2_X$ values. Note that $A2_X$ is typically consistent with zero within the uncertainties, and in principle it could be assumed to be equal to zero.

   In general, we define the total dust-corrected metallicity of a metal $X$ as
    \begin{equation} \label{eq:meth2}
       [X/\mathrm{H}]_{\mathrm{tot}} = [X/\mathrm{H}] - \delta_X, 
    \end{equation}
    where [$X$/H]$_{\mathrm{tot}}$ represents the total metallicity of the neutral ISM, including dust and gas, and $[X/\mathrm{H}]$ are the observed gas-phase abundances. For $i$ individual gas components along a line of sight, we combine these two equations to obtain 
    \begin{align}
        y_i =& a_i + B2_X \times \mathrm {[Zn/Fe]}_{i, \mathrm{fit}}, \label{eq:y_i=ai+depl}
    \end{align}
        which is a linear relation, $y_i = a_i + bx$, where 
    \begin{align}
        y_i =& \log{N(X)_i} - X_{\odot} + 12 - A2_X , \label{eq:yi} \\
        a_i =& [\mathrm{M/H}]_{i} + \log{N(\mathrm{H})_i.} \label{eq:ai}
    \end{align}
        Here, $\log{N(X)_i}$, [M/H]$_i$ and $\log{N(\mathrm{H})_i}$ are the metal column density, dust corrected metallicity and total hydrogen (i.e. H\,\textsc{i + H$_2$, ignoring the ionised hydrogen H\,\textsc{ii})} column density for individual gas components along the line of sight. $y_i$ represents the `normalised' (by its Solar abundance) metal column densities. 
        
        We plot the $y_i$ values for each metal against its corresponding $B2_X$, which constitutes what we call a `metal pattern' \citep[or `depletion pattern' in][]{Ramburuth-Hurt+2023} In the case where there is no dust depletion (or any other causes of deviation), $y_i$ for each metal would be similar. If there is only dust depletion at play, we would expect the $y_i$ values to align on a straight line with a gradient [Zn/Fe]$_{\mathrm{fit, }i}$, which represents the overall strength of depletion. The $y$-intercept of this straight line, $a_i = \mathrm{[M/H]}_i + \log{N(\mathrm{H})_i}$, is the sum of the total hydrogen column density and dust-corrected metallicity of the individual gas component, and it is degenerate. To demonstrate this degeneracy, consider a line of sight with two individual components and the following two (of several possible) scenarios. If we have $a_{1} < a_{2}$, it is possible for component 2 to either have the majority of the hydrogen and a low metallicity or it could have less hydrogen but a higher metallicity than component 1. 
        
        With the relative method, we characterise the level of dust depletion for individual gas components along the line of sight regardless of its hydrogen content. Deviations from the fitted linear relation are possible and could be due to processes such as ionisation or nucleosynthesis (e.g. $\alpha$-element enhancement), or peculiar depletion due to local physical conditions of a gas cloud. Potential peculiarities of dust depletion were observed for the Magellanic Clouds \citep{Jenkins&Wallerstein2017}, but are likely due to $\alpha$-element enhancement \citep{Konstantopoulou+2022, DeCia+2024}.  The relative method is able to distinguish between dust depletion, $\alpha$-element enhancement and metallicity variations.

   An alternative to the relative method is the $F^*$ method \citep{Jenkins2009}. We do not discuss this method in this paper, but we not that the values for depletion we determine here are related to the $F^*$ method by the equation $F^* = 1.05 \times$ [Zn/Fe]$_{\mathrm{fit}} - 0.86$ \citep[based on data from][]{DeCia+2021}. 

   While the metallicity of components is not possible to constrain, the dust depletion of each component is well-constrained by Equation \ref{eq:y_i=ai+depl}. Metal patterns give us information on any variation in the amount of dust between gas clouds along a line of sight. We are able to study the dust depletion of the ISM on a component-by-component basis without making assumptions on the metallicity or the hydrogen distribution. 

    \begin{table}
        \caption{Coefficients of the depletion sequences for metals $X$ \citep{Konstantopoulou+2022}. The values for P are those from \cite{Konstantopoulou+2023}.}            
        \label{table:depletion sequences}     
        \centering                         
        \begin{tabular}{c | c c}        
            \hline\hline                 
            $\delta_X$ & $A2_X$ & $B2_X$ \\   
            \hline                        % in
            $\delta_{\mathrm{Zn}}$ & 0.00 $\pm$ 0.01    & $-0.27$ $\pm$ 0.03 \\
            $\delta_{\mathrm{O}}$  & 0.00 $\pm$ 0.00    & $-0.20$ $\pm$ 0.05 \\
            $\delta_{\mathrm{P}}$  & 0.08 $\pm$ 0.05    & $-0.26$ $\pm$ 0.08  \\
            $\delta_{\mathrm{S}}$  & 0.01 $\pm$ 0.02    & $-0.48$ $\pm$ 0.04  \\
            $\delta_{\mathrm{Si}}$ & $-0.04$ $\pm$ 0.02 & $-0.75$ $\pm$ 0.03 \\ 
            $\delta_{\mathrm{Mg}}$ & 0.01 $\pm$ 0.03    & $-0.66$ $\pm$ 0.04 \\
            $\delta_{\mathrm{Ge}}$ &0.00       & $-0.40$ $\pm$ 0.04 \\
            $\delta_{\mathrm{Mn}}$ & 0.07 $\pm$ 0.02    & $-1.03$ $\pm$ 0.03 \\ 
            $\delta_{\mathrm{Cr}}$ & 0.12 $\pm$ 0.01    & $-1.30$ $\pm$ 0.01 \\ 
            $\delta_{\mathrm{Fe}}$ & $-0.01$ $\pm$ 0.03 & $-1.26$ $\pm$ 0.04 \\ 
            $\delta_{\mathrm{Ti}}$ & $-0.07$ $\pm$ 0.03 & $-1.67$ $\pm$ 0.04 \\ 
             
            \hline                                   
        \end{tabular}
    \end{table}

% -------------------------------------------------------------------------

\subsection{Simulating clouds: exploring the metallicities of individual gas components} \label{sec:simulating clouds}

It can be argued that the distribution of \ion{O}{I} can be used to give some insight into the distribution of hydrogen, both because it traces \ion{H}{I} (due to their similar ionisation potential), and because it is not highly depleted into dust. The caveat with using the \ion{O}{I} gas fraction as a direct tracer of the hydrogen fraction is that it makes the assumption that the metallicity of each gas component is the same. We choose not to do this here in order to avoid this assumption.

    The distinctive difference between metal patterns described here and the more typical abundance patterns (the study of the observed abundances of elements with different refractory properties) is that we do not know the hydrogen column densities for individual components. Therefore, metals patterns cannot directly provide the metallicity of the component.

    In order to investigate possible metallicities for components along our lines of sight, we perform simulations. The simulations take into account the observed column densities and then explore possible hydrogen gas distributions between the components. In practice, we use the $y$-intercept $a_i$ derived from the linear fit to the metal patterns, and calculate the metallicity based on different combinations of hydrogen gas fractions. Rewriting Equation \ref{eq:ai} gives
    \begin{align} \label{eq. calculate metallicity}
        \mathrm{[M/H]}_i = a_i - \log{\big{(}~ f_i \times N(\mathrm{H})_{\mathrm{tot}}~ \big{)}}.
    \end{align}
    Here, $N(\mathrm{H})_{\mathrm{tot}}$ is the total hydrogen column density along the full line of sight, which can also be written in terms of the gas fraction:
    \begin{align} \label{eq. fi gas fractions}
    N(\mathrm{H})_i = f_i  \times N(\mathrm{H})_{\mathrm{tot}}, 
    \end{align}
    such that $\sum_{i} f_i = 1$.

\begin{figure}
        \centering
        \includegraphics[width=0.4\textwidth]{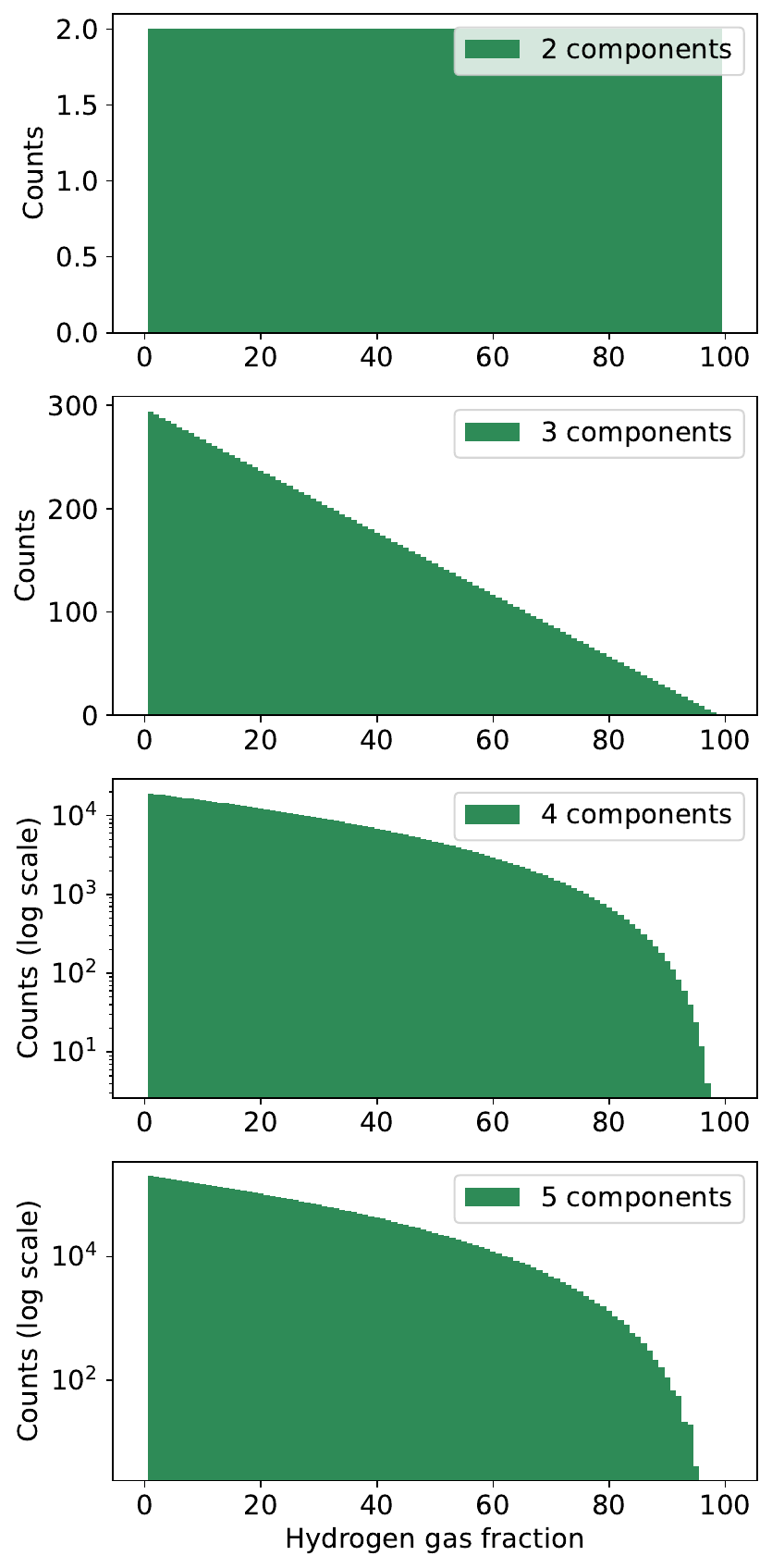}
        \caption{The distributions of total hydrogen gas fractions that we use to infer the metallicities of individual components. From the top to the bottom panel are the distributions for 2 to 5 components respectively. The histograms for 4 and 5 components are plotted in log-scale for visual clarity.}
        \label{fig:HI-gas-fractions}
\end{figure}

    The hydrogen gas fraction combinations are generated as follows. In general, for $m$ number of whole integers that add up to 100, the number of combinations is $ = \binom{100-1}{m-1} = \frac{99!}{(m-1)! 100!}$. Here, we first generate all these possible permutations for each number of components, i.e. $m = 2, 3, 4, 5$. We then generate all the permutations of these and remove any duplicates. This process produces over three million combinations for $m=5$, therefore, to optimise subsequent calculations, we randomly sample 1~000~000 combinations from these, where 1~000~000 is chosen because it yields enough combinations to cover the parameter space between 0\% -- 100\% in intervals of 1\%. For 2 to 4 components, we obtain 99, 4~851, 156~849, combinations respectively, which is visualised in Fig. \ref{fig:HI-gas-fractions}. The aim of this procedure is to have combinations that span the parameter space in intervals of 1\%.

    Then, for each gas fraction combination, we calculate the corresponding set of metallicities from Equation \ref{eq. calculate metallicity}, which we also refer to as a realisation. We do this for each combination of gas fractions and for each target. This gives us the possible ranges of metallicities for the components along each line of sight. In principle, each realisation is equally likely.
    
    To investigate the minimum possible range in metallicity along each line of sight, we compute the difference between the maximum and minimum metallicity for each realisation, and extract the realisation that produces the minimum difference. We interpret this is as the minimum possible variation in metallicity.
    
    We test this methodology by combining our observations towards HD~206267 with a sight line towards the SMC, which has known metallicity and $\log{N(\ion{H}{I})}$ from \cite{DeCia+2024}, to create a synthetic sight line. The choice of an SMC sight line is to demonstrate that our method can distinguish between the low metallicity of the SMC and the higher metallicity of the Milky Way. The test confirms that we are able to recover a 0.6 dex difference in metallicity between both the Milky Way components and the SMC (i.e. [M/H]$_{\mathrm{MW 1}}$ -- [M/H]$_{\mathrm{SMC}}$ > 0.6 and [M/H]$_{\mathrm{MW 2}}$ - [M/H]$_{\mathrm{SMC}}$ > 0.6) for 5 out of 4865 most-likely realisations. These 5 realisations also recover the metallicity and hydrogen gas fraction of the SMC cloud. The complete test is detailed in Appendix \ref{appendix:method-test}.

%======================================================================
%======================================================================
%======================================================================

\section{Results and discussion} \label{sec:results}
   
   From the analysis of the metal patterns, we derive the dust depletion of individual components and calculate the range in dust depletion along each of the eight lines of sight as the difference between the maximum and minimum depletion. We generally do not see strong evidence of deviations from the linear behaviour of the metal patterns of the individual components in this sample, and therefore assume that the relative method can be applied here. One exception to this may be the case of $\theta^1$ Ori C, where a potential overabundance of the iron-group elements or under-abundance of \ion{Ti{II}} is observed. 
   
   We find ranges up to 1.19 dex in one case and present these results in Section \ref{subsec: variations in dust depletion }. Using simulations, we further explore the possible metallicities and gas fractions of the individual components, which can be constrained to 0.1 dex for some lines of sight. The simulations results are presented in Section \ref{subsec: variations in metallicity}. We analyse the realisation that produces the minimum range in metallicity for each sight line in Section \ref{subsec: minimum variation in metallicity}. In Section \ref{subsec: abundance pattern plots} we present the abundance patterns along the full lines of sight. 
   
% --------------------------------------------------------------------
    
    \subsection{Variations in dust depletion} \label{subsec: variations in dust depletion }
    
\begin{figure}[htbp]
    \centering
    \includegraphics[width=0.4\textwidth]{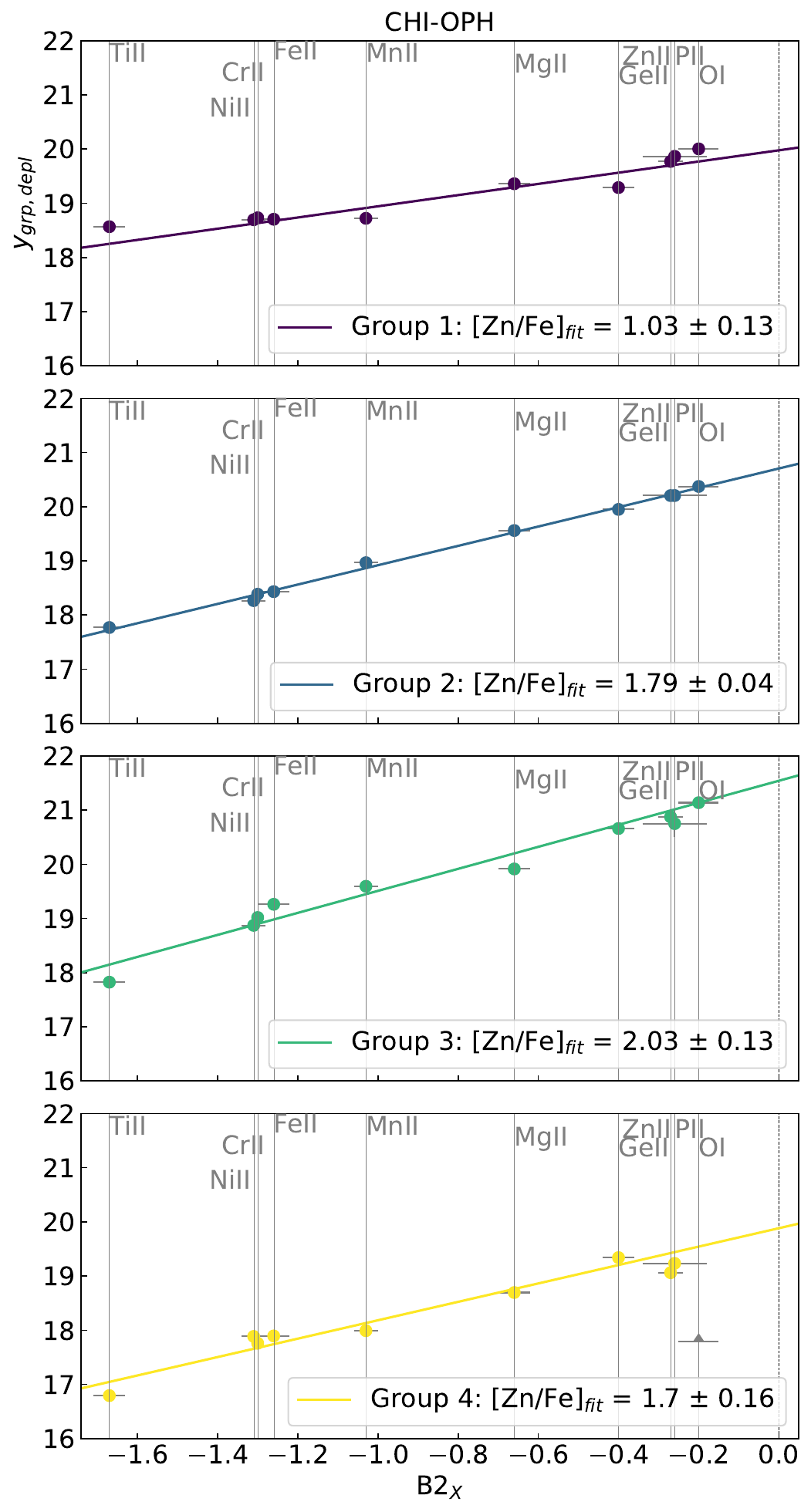}
    \caption{Metal patterns for grouped components towards $\chi$ Oph. This line of sight has the highest value for dust depletion out of our sample, with [Zn/Fe]$_{\mathrm{fit, 3}} =  2.03 \pm 0.13$ dex.}
    \label{fig:chi-oph_metal-patterns}
\end{figure}

    In our component-by-component analysis, some components show very high levels of dust depletion. The highest level of dust depletion we find is [Zn/Fe]$_{\mathrm{fit, }i} = $ 2.03 $\pm$ 0.03 dex (or $F^*$ = 1.5) for group 3 in $\chi$ Oph. This value is larger than the maximum dust depletion reported by \cite{DeCia+2021}, which is 1.32 dex, observed towards $\iota$ Ori. Here we show the metal pattern for $\chi$ Oph in Fig. \ref{fig:chi-oph_metal-patterns}. In Table \ref{tbl:metal-patterns-results-all-targets} we present [Zn/Fe]$_{\mathrm{fit, }i}$ and $a_i$ for each component along each sight line. The metal patterns for the remaining targets are presented in Appendix \href{https://zenodo.org/records/14793736}{D}. 
    
\begin{figure}
        \centering
        \includegraphics[width=0.4\textwidth]{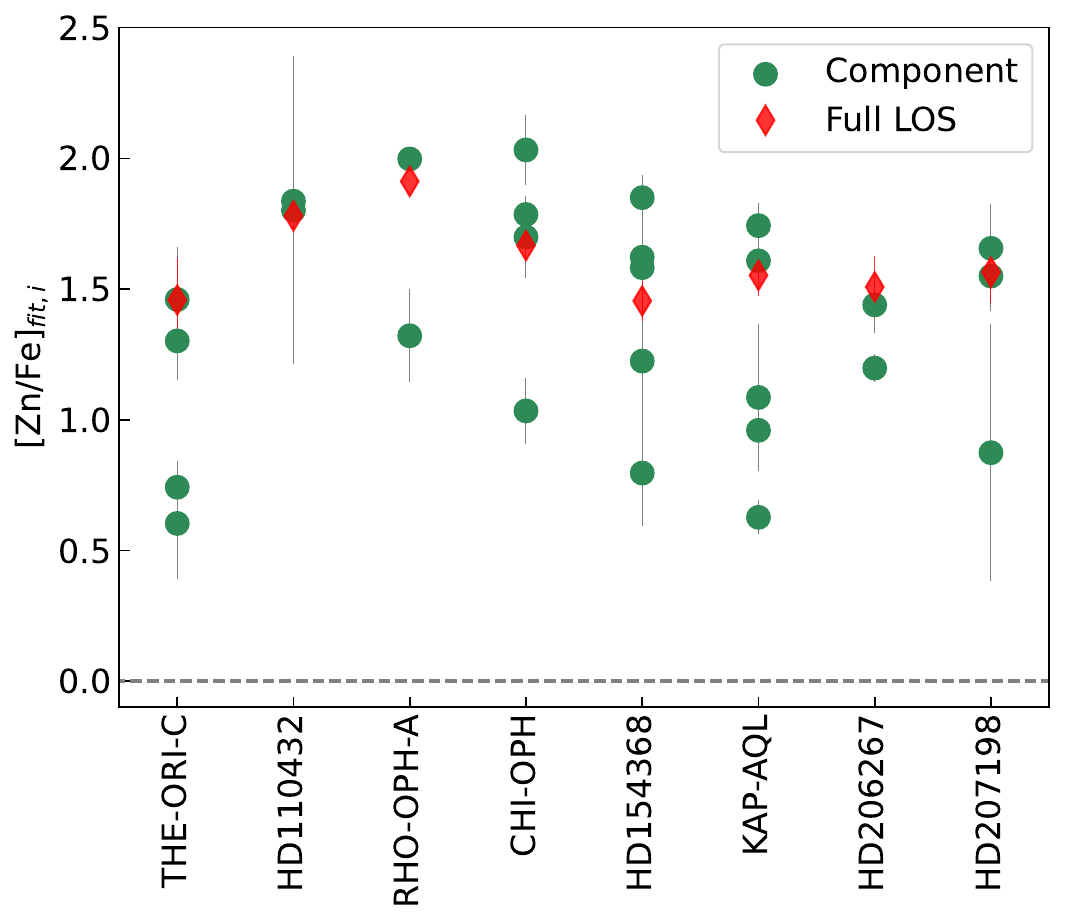}
        \caption{Depletion factors for individual components along each line of sight are represented by green points. The [Zn/Fe]$_{\mathrm{fit}}$ values for the entire line of sight are plotted as red diamonds.}
        \label{fig:depletion-all-targets}
\end{figure}

    We also observe a large range in dust depletion along sight lines. Fig. \ref{fig:depletion-all-targets} shows that, for four of our sight lines, the difference between maximum and minimum [Zn/Fe]$_{\mathrm{fit, }i}$ values reaches $\sim$ 1.0 dex. For most targets, several components have significantly higher or lower levels of dust depletion compared to that of the total sight line. This result confirms that total line-of-sight analyses \citep[e.g.][]{DeCia+2021, Ritchey+2023} can wash out the level of diversity in depletion along sight lines. Indeed, it is consistent with established variations in [Na/Ca] in many Milky Way sight lines \citep{RoutlySpitzer1957, SilukSilk1974} and with the few studies of individual components \citep{Fitzpatrick&Spitzer, Welty+2020_HD62542}.


\begin{figure}
        \centering
        \includegraphics[width=0.4\textwidth]{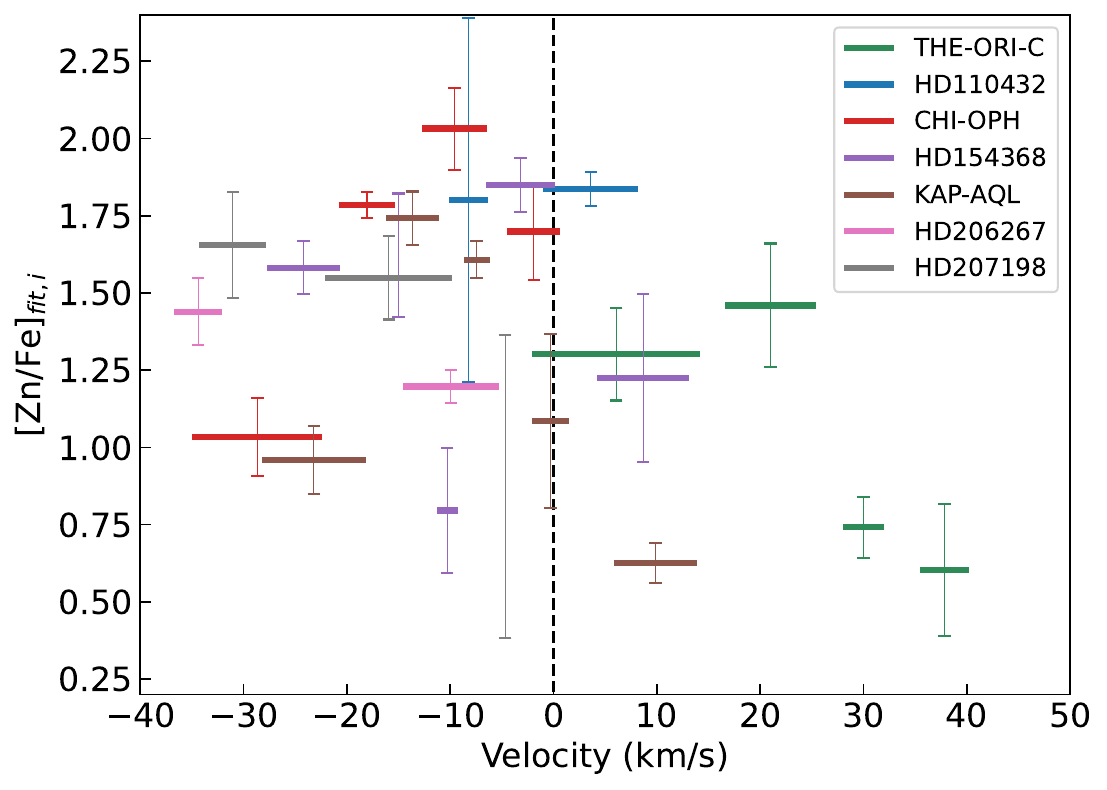}
        \caption{Velocity-depletion plot for groups of components.}
        \label{fig:velocity-depletion}
\end{figure}

    In Fig. \ref{fig:velocity-depletion} we plot the velocity ranges of the groups of components against their depletion. This is to investigate, firstly, any relationship between kinematics and the depletion, and secondly, any trends that could be associated with the Local Bubble \citep{Lallement+2003, Zucker+2022}. The Local Bubble is a pocket of plasma with low density and high temperature in which the Sun sits. Outside of the Bubble is a shell of cold, neutral gas. \cite{Zucker+2022} find that several of the nearby star-forming regions sit on the edge of the Local Bubble, at a velocity of 6.7 $\pm$ 0.45 km~s$^{-1}$, like the Orion region. We do not find any significant relationship between velocity and depletion, or any patterns, for example systematic $\alpha$-element enhancement of components around 6.7 km~s$^{-1}$, that could be associated with the Local Bubble. Fig. \ref{fig:velocity-depletion} agrees well, however, with the velocity-depletion results of \cite{RoutlySpitzer1957} and \cite{SilukSilk1974}. There may be a slight anti-correlation between absolute velocity and the level of dust depletion in these works, although we remain uncertain about the potential causes. \cite{RoutlySpitzer1957, SilukSilk1974} interpret this as evidence for the partial destruction of dust grains within velocities of $\sim 40$ km~s$^{-1}$, possibly due to SNe shocks. 

\begin{table}
\centering
\caption{Results from the metal patterns. The values in square brackets are based on a fit to only two points, and are therefore not reliable.}
\label{tbl:metal-patterns-results-all-targets}
\begin{tabular}{c|c|c|c}
\hline \hline
Target &  Group & [Zn/Fe]$_{i,\mathrm{fit}}$ & a$_{i, \mathrm{depl}}$ \\
\hline
 $\theta^1$ Ori C &    1 &             1.30 $\pm$ 0.15 &       20.07 $\pm$ 0.12 \\
                &               2 &            1.46 $\pm$ 0.20 &       21.65 $\pm$ 0.22 \\
                &               3 &             0.74 $\pm$ 0.10&        19.26 $\pm$ 0.10 \\
                &               4 &             0.60 $\pm$ 0.21 &       18.67 $\pm$ 0.29 \\
\hline
      HD~110432 &               1 &             1.80 $\pm$ 0.59 &       20.15 $\pm$ 0.73 \\
                &               2 &            1.84 $\pm$ 0.06 &       21.57 $\pm$ 0.07 \\
\hline
   $\rho$ Oph A &               1 &            1.32 $\pm$ 0.18 &       20.15 $\pm$ 0.19 \\
                &               2 &             [2.0]         &        [22.02]         \\
\hline
     $\chi$ Oph &               1 &            1.03 $\pm$ 0.13 &       19.98 $\pm$ 0.12 \\
                &               2 &            1.79 $\pm$ 0.04 &       20.71 $\pm$ 0.04 \\
                &               3 &            2.03 $\pm$ 0.13 &       21.58 $\pm$ 0.18 \\
                &               4 &             1.70 $\pm$ 0.16 &        20.02 $\pm$ 0.20 \\
\hline
      HD~154368 &               1 &            1.58 $\pm$ 0.08 &       20.44 $\pm$ 0.08 \\
                &               2 &             1.62 $\pm$ 0.20&       20.27 $\pm$ 0.23 \\
                &               3 &              0.80 $\pm$ 0.20&       20.18 $\pm$ 0.25 \\
                &               4 &            1.85 $\pm$ 0.09 &         21.60 $\pm$ 0.10 \\
                &               5 &            1.22 $\pm$ 0.27 &       20.04 $\pm$ 0.35 \\
\hline
    $\kappa$ Aql&               1 &            0.96 $\pm$ 0.11 &        19.6 $\pm$ 0.11 \\
                &               2 &            1.74 $\pm$ 0.09 &       21.16 $\pm$ 0.01 \\
                &               3 &            1.60 $\pm$ 0.06 &        20.56 $\pm$ 0.06 \\
                &               4 &            1.09 $\pm$ 0.28 &       19.32 $\pm$ 0.34 \\
                &               5 &            0.63 $\pm$ 0.06 &       19.33 $\pm$ 0.07 \\
\hline
      HD~206267 &               1 &            1.41 $\pm$ 0.07 &       20.01 $\pm$ 0.08 \\
                &               2 &             1.2 $\pm$ 0.05 &       21.26 $\pm$ 0.07 \\
\hline
      HD~207198 &               1 &            1.25 $\pm$ 0.25 &        20.7 $\pm$ 0.29 \\
                &               2 &            1.55 $\pm$ 0.14 &       21.49 $\pm$ 0.17 \\
                &               3 &            0.87 $\pm$ 0.49 &       19.71 $\pm$ 0.68 \\

\hline
\end{tabular}
\end{table}
        
% -------------------------------------------------------------------- 

\subsection{Variations in metallicity} \label{subsec: variations in metallicity}

\begin{figure}
    \centering
    \includegraphics[width=0.4\textwidth]{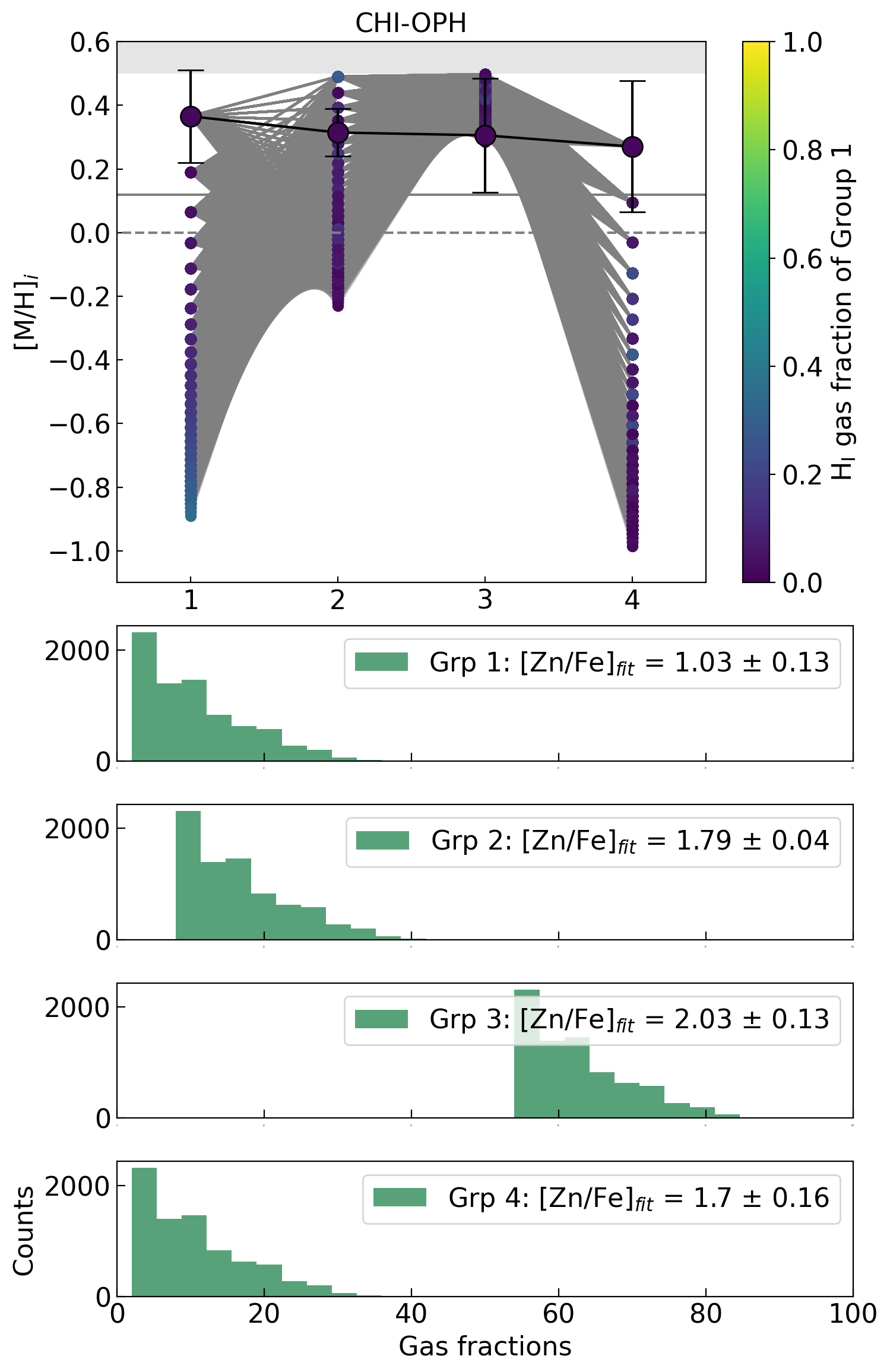}
    \caption{The simulation realisations for the most-likely cases (where the metallicities for all groups are $\leq 0.5$ dex) for $\chi$ Oph. The solid grey horizontal line in the top panel represents the metallicity along the whole line of sight, taking into account all of the metals. The points with black edge colours connected by the black line highlight the case with the minimum differences in metallicity. The grey-shaded region above 0.5 dex shows the upper bound criteria for the metallicities. The colours of the points correspond to the gas fraction of the first group $N(\mathrm{H}_1)$ for each realisation (i.e. approximately $1\% \leq f_1 \leq 45\%$), and is represented by the colour bar. The histograms in the following panels are the total hydrogen gas fractions, $f_i$ for the most-likely cases. We include the level of dust depletion [Zn/Fe]$_{\mathrm{fit, i}}$ in the legends.}
    \label{fig:most-likely-met+HI-CHI-OPH}
\end{figure}

The simulations we perform produce thousands of possible metallicity combinations for individual gas components based on hydrogen gas fraction combinations, as described in Section \ref{sec:simulating clouds}. From these simulations we obtain the possible ranges of metallicities for each component (see Fig. \ref{fig:most-likely-met+HI-CHI-OPH} for $\chi$ Oph results). In principle, all of the realisations are equally possible because we do not state any preference for particular hydrogen gas fraction combinations. However, we do make a physically motivated upper cut on possible metallicities, which narrows the range of their possible values. The threshold is based on the maximum metallicity found in stars in the Milky Way's disk, which \cite{Nepal+2024_MW-metallicity} find to be 0.4 dex. To be conservative, we assume that the upper limit on the gas-phase metallicity could be slightly higher and therefore impose a limit of [M/H]$_i \leq$ 0.5 dex. This limit decreases the range of allowed metallicities for most components, and can even constrain the possible metallicity of a component down to within an interval of $\sim$ 0.3 dex including uncertainties, for example group 3 of $\chi$ Oph and group 2 of HD~110432. We show the allowed metallicities for each component along our lines of sight in Appendix \href{https://zenodo.org/records/14793736}{E} (Figures E.1 -- E.5). The highlighted points in these figures indicate the realisation that produces the smallest range in metallicity. This is discussed later in Section \ref{subsec: minimum variation in metallicity}.

In five lines of sight ($\theta^1$ Ori C, $\chi$ Oph, HD~154368, $\kappa$ Aql, HD~207198), we find that the component with the highest [Zn/Fe]$_{\mathrm{fit, }i}$ holds the majority of the hydrogen gas, and also likley has super-Solar metallicity. These five cases and these results are listed in Table \ref{table:sim-metallicities}. The exceptions are $\rho$ Oph A, for which we cannot perform this analysis due to limited data, and HD~110432 and HD~206267, which both have two components of the same [Zn/Fe]$_{\mathrm{fit, }i}$ within the uncertainties. Below we further discuss the results of the most-likely realisations for some of the targets. 

\begin{table*}
\centering
\caption{Five sight lines where the component with the highest depletion also contains the majority of the hydrogen gas. For $\chi$ Oph and $\kappa$ Aql, we find that the metallicities of these components are also constrained to super-Solar.}
\label{table:sim-metallicities}
\begin{tabular}{c|c|c|c|c}
\hline \hline
Target &  Group & [Zn/Fe]$_{\mathrm{fit, i}}$ & $f_i$ & [M/H]$_i \geq $  \\
\hline
$\theta^1$ Ori C    & 2 & 1.46 $\pm$ 0.20    & 37\% $< f_2 <$ 97\% & 0.08 $\pm$ 0.31 \\
$\chi$ Oph    & 3 & 2.03 $\pm$ 0.13   & 54\% $< f_3 <$ 88\% & 0.29 $\pm$ 0.21 \\
HD~154368     & 4 & 1.85 $\pm$ 0.09            & 38\% $< f_4 <$ 89\% & 0.13 $\pm$ 0.25 \\
$\kappa$ Aql  & 2 & 1.74 $\pm$ 0.09   & 43\% $< f_2 <$ 84\% & 0.20 $\pm$ 0.15 \\
HD~207198     & 2 & 1.55 $\pm$ 0.14            & 39\% $< f_2 <$ 77\% & 0.20 $\pm$ 0.21 \\
\hline
\end{tabular}
\end{table*}

\begin{figure}
    \centering
    \includegraphics[width=0.4\textwidth]{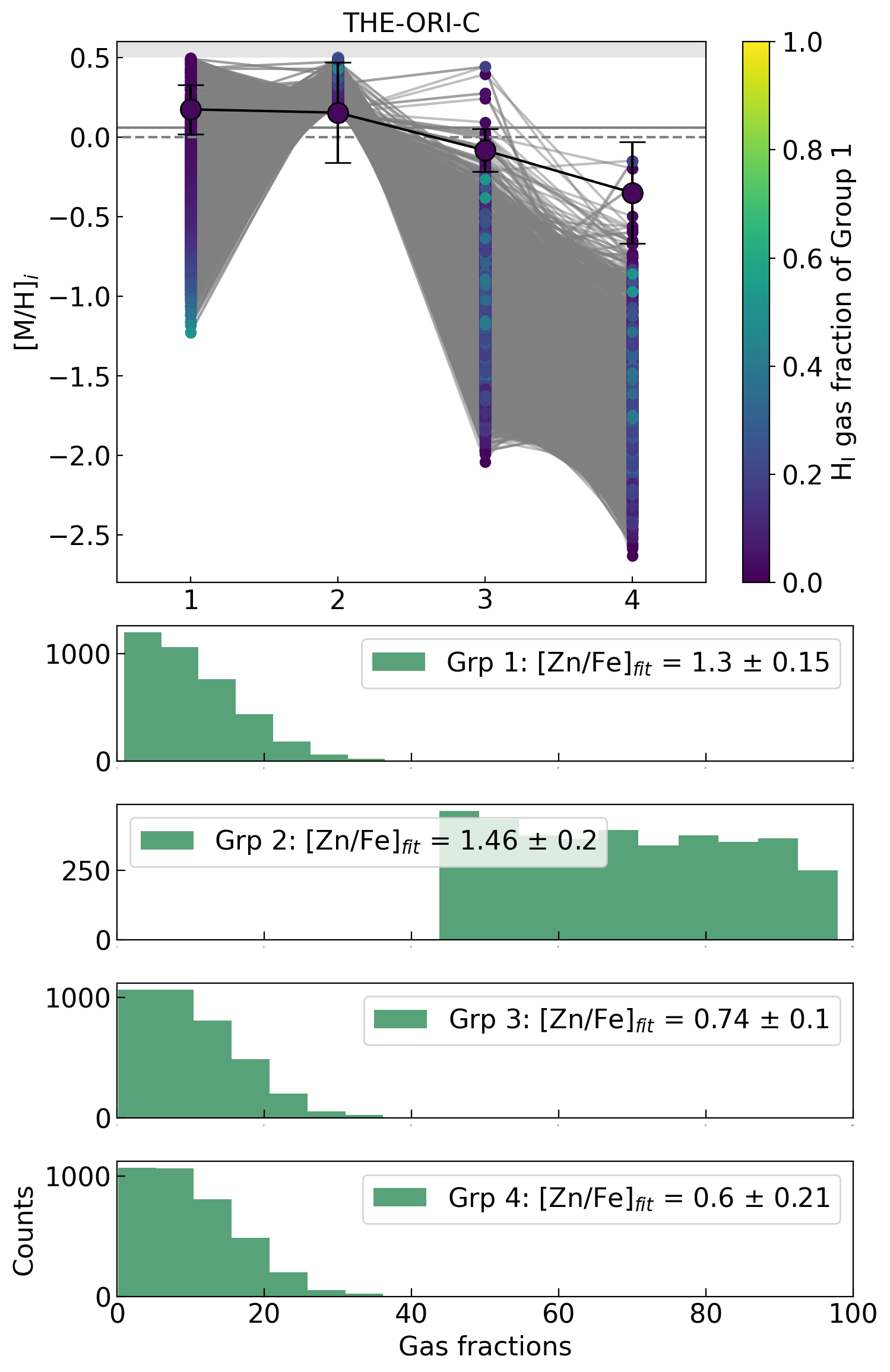}
    \caption{The same as Fig. \ref{fig:most-likely-met+HI-CHI-OPH}, but for $\theta^1$ Ori C.}
    \label{fig:most-likely-met+HI-THE-ORI-C}
\end{figure}

\subsubsection{$\theta^1$ Ori C} \label{sec:metallicity Theta ori c}
In addition to group 2 having high depletion, this component has super-Solar metallicity and the bulk of the hydrogen gas (see Fig. \ref{fig:most-likely-met+HI-THE-ORI-C}). Additionally, there is evidence that this gas cloud is attributed to the gas Orion nebula itself \citep{Price+2001}. We also find that group 4 has the lowest amount of dust depletion, sub-Solar metallicity, and $\leq 40\%$ of the hydrogen gas.  

\subsubsection{HD~110432}
The two components along this line of sight have similar depletion: [Zn/Fe]$_{\mathrm{fit, 1}} = 1.80 \pm 0.59$ and [Zn/Fe]$_{\mathrm{fit, 2}} = 1.84 \pm 0.06$ (see Fig. \href{https://zenodo.org/records/14793736}{E.1}). Although the metallicity of group 2 can be constrained to be $\geq 0.40 \pm 0.19$ dex, we cannot constrain the metallicity for group 1 to the same degree. Our simulations show that it is possible for the two components to have similar depletion but have vastly different metallicities. For example, one realisation for these two components produces [M/H]$_1 = -0.39 \pm 0.12$ and [M/H]$_2 = 0.50 \pm 0.12$, which is a difference of $\sim 0.89$ dex. 

\subsubsection{$\rho$ Oph A}
It is not possible to draw any conclusions about metallicity variations along this line of sight due to lack of data.

\subsubsection{$\chi$ Oph}
The simulations for this line of sight constrain the gas fractions for each group within a range of $\sim$ 30\%, with group 3 having the majority of the gas (55\% -- 85\%, see Fig. \ref{fig:most-likely-met+HI-CHI-OPH}). This line of sight has the component with the highest depletion of all our targets: [Zn/Fe]$_{\mathrm{fit, 4}} = 2.03 \pm 0.13$, which we also constrain to have super-Solar metallicity [M/H]$_4 \gtrsim$ 0.3.

\subsubsection{HD~206267}
The groups along this line of sight appear to be similar in terms of depletion, gas fractions and metallicity (see Fig. \href{https://zenodo.org/records/14793736}{E.4}). Gas fractions overlap significantly, making it more difficult to disentangle the degeneracy between metallicity and hydrogen gas fraction. It is possible that the gas we probe along this line of sight is chemically homogeneous, also considering that we found the minimum difference in metallicities to be zero.

\subsection{Minimum differences in metallicity} \label{subsec: minimum variation in metallicity}

Here we report and analyse the realisations that produce the smallest difference in metallicity for each line of sight. As described in Section \ref{sec:simulating clouds}, this realisation is as equally likely as each other realisation in principle. We interpret it as the minimum possible level of variation in metallicity. The metallicities and gas fractions of these realisations are tabulated in Table \ref{tbl:minimum-variation-results} for all sight lines. The uncertainties here are propagated from the fits to the metal patterns and the uncertainties in $N(\mathrm{H})_{\mathrm{tot}}$. In Figures \ref{fig:most-likely-met+HI-CHI-OPH} and \href{https://zenodo.org/records/14793736}{E.1} -- \href{https://zenodo.org/records/14793736}{E.5} these `minimum-difference' realisations are highlighted by the larger points and joined with a black solid line.  

For HD~110432 and HD~206267, both with two components, we find the difference in metallicity to be zero. For the remaining lines of sight, the difference in metallicity is between 0.02 and 1.02 dex. 

For $\chi$ Oph, HD~110432 and HD~154368 we find that the metallicities in the minimum-difference case are all higher than that for the total line of sight (indicated by the solid grey line). While the likelihood is, in principle, the same as others, this discrepancy may suggest that these are not highly likely options.

\begin{figure}
    \centering
    \includegraphics[width=0.4\textwidth]{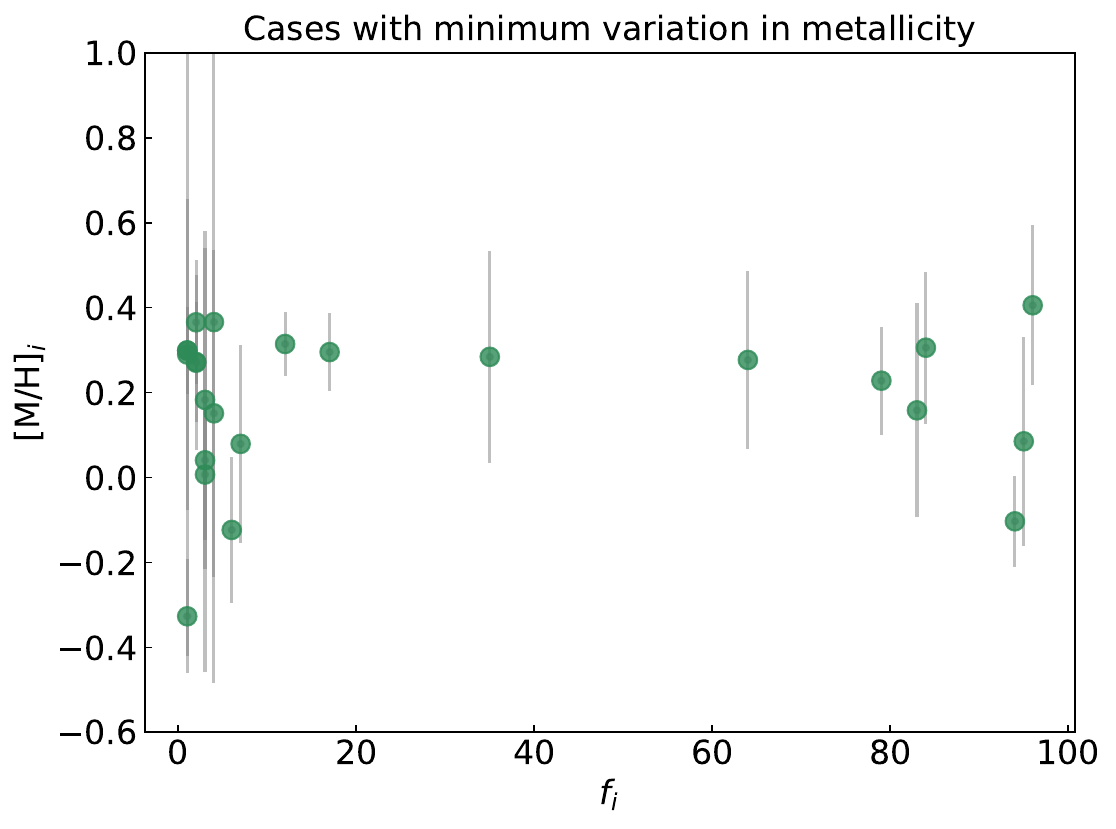}
    \caption{Plot of metallicity realisations for individual components as a function of their gas fraction for the cases with minimum difference in metallicity. Here we see that these cases require the gas fractions to be either very high or very low.}
    \label{fig:min-variation-case_gasfrac-vs-met}
\end{figure}

We find that the gas fractions for the minimum-difference realisations are on the extreme ends, only below $\sim 20\%$ or above $\sim 70\%$, shown in Fig. \ref{fig:min-variation-case_gasfrac-vs-met}. This means that, in order to have $\sim$ the same metallicity in all components along the line of sight, one of the components has to carry the vast majority of the hydrogen gas. 

We perform a statistical $z$-test to quantify how significant the difference in metallicity is when considering the uncertainties:
\begin{equation}
z = \frac{(\mathrm{[M/H]_{max}} - \mathrm{[M/H]_{min}})}{\sqrt{\Delta \mathrm{[M/H]_{min}} ^2 + \Delta \mathrm{[M/H]_{max}}^2}},
\end{equation}
where $\mathrm{[M/H]_{max}}$ and $\mathrm{[M/H]_{min}}$ are the maximum and minimum metallicities, and $\Delta \mathrm{[M/H]_{max}}$ and $\Delta \mathrm{[M/H]_{min}}$ are their respective uncertainties. A $z$-value above 3 means that the difference between the maximum and minimum metallicities is statistically significant to the 3$\sigma$ level. 

\begin{figure}
        \centering
        \includegraphics[width=0.4\textwidth]{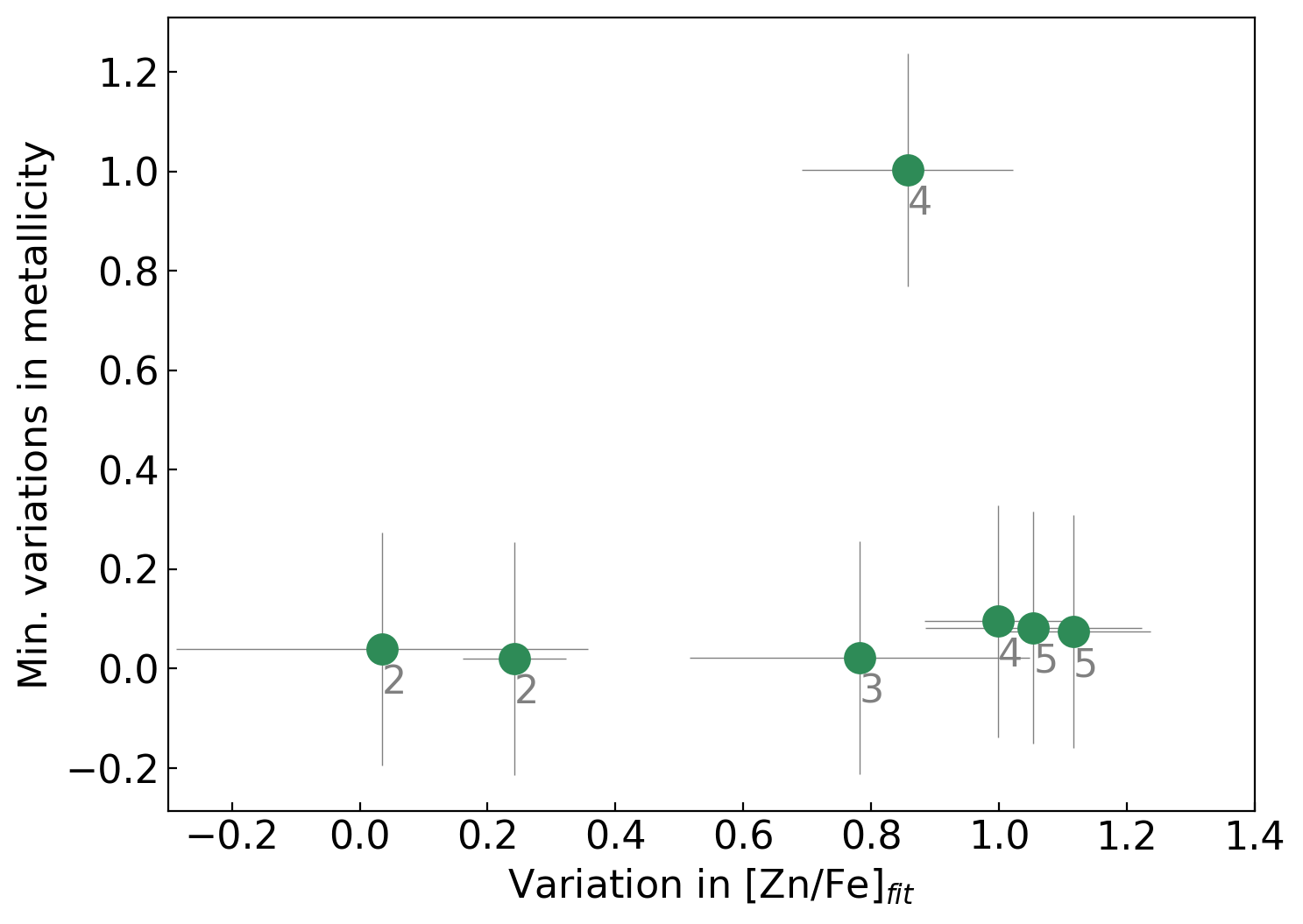}
        \caption{Variation in [Zn/Fe]$_{\mathrm{fit}}$ plotted against the minimum difference in metallicity for all targets. The grey numbers indicate the number of groups of components along the line of sight.}
        \label{fig:min-variations-ZnFe-variations}
\end{figure}

The line of sight towards $\theta^1$ Ori C has the most statistically significant difference in metallicity, although it is still small with only 2.5$\sigma$. The results from this exercise are tabulated in Table \ref{table:min-met-var}. In Fig. \ref{fig:min-variations-ZnFe-variations}, we plot these minimum differences in metallicity against the difference in dust depletion [Zn/Fe]$_{\mathrm{fit}}$, and do not find any strong correlation.

\begin{figure}
        \centering
        \includegraphics[width=0.5\textwidth]{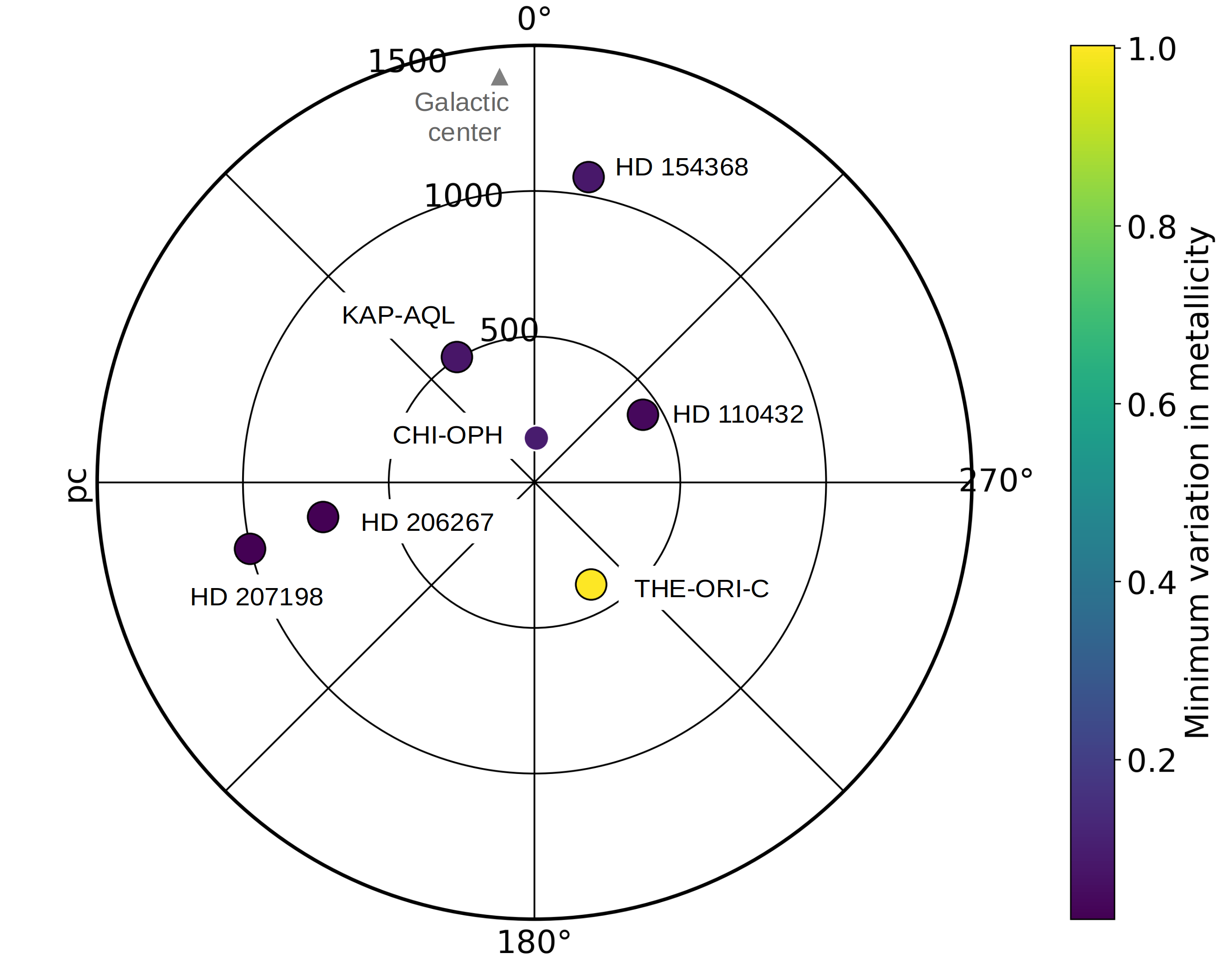}
        \caption{Labelled target map, colour-coded by the minimum differences in metallicity based on our realisations. The Sun is positioned at the center.}
        \label{fig:target-map}
\end{figure}

Fig. \ref{fig:target-map} shows the minimum differences in metallicity for our lines of sight compared to the positions of the targets with respect to the Sun. This is to investigate whether the length of the line of sight (i.e. the distance from Earth to the target star) has an impact on the amount of variation in metallicity. We do not find any coherent spatial variation linked to the metallicity variations.

It is important to point out that small shifts in the gas fractions can produce drastically different ranges in metallicity and result in the metallicities. For example, the minimum difference between the two groups of components for HD~110432 gives gas fractions of $f_1 = 4\%$ and $f_2 = 96\%$ and metallicities of [M/H]$_1 = 0.37 \pm 0.85$ and [M/H]$_2 = 0.41 \pm 0.19$. However, if we move $\sim 10\%$ of the gas from component 2 to component 1 so that they become $f_1 = 13\%$ and $f_2 = 86\%$, then the range in metallicity increases to 0.6 dex with [M/H]$_1 = -0.15 \pm 0.85$ and [M/H]$_2 = 0.45 \pm 0.19$. 

\begin{table}
\centering
\caption{The metallicities and corresponding gas fractions for the cases with the minimum difference in metallicity.}
\label{tbl:minimum-variation-results}
\begin{tabular}{c|c|c|c}
\hline \hline
Target &  Group &  [M/H]$_i$ & $f_i$ (\%) \\
\hline
 $\theta^1$ Ori C &    1 & 0.00    $\pm$ 0.15 & 3 \\
                &               2 & 0.10    $\pm$ 0.31 & 95 \\
                &               3 & $-$0.33 $\pm$ 0.13 & 1 \\
                &               4 & $-$0.92 $\pm$ 0.32 & 1 \\
\hline
      HD~110432 &               1 & 0.37 $\pm$ 0.85 &  4 \\
                &               2 & 0.41 $\pm$ 0.19 & 96 \\
\hline
     $\chi$ Oph &               1 &  0.37 $\pm$ 0.15 &  2 \\
                &               2 & 0.31 $\pm$ 0.07 & 12 \\
                &               3 & 0.31 $\pm$ 0.18 & 84 \\
                &               4 & 0.27 $\pm$ 0.21 &  2 \\
\hline
      HD~154368 &               1 & 0.22 $\pm$ 0.23 &  5 \\
                &               2 & 0.15 $\pm$ 0.38 &  4 \\
                &               3 & 0.18 $\pm$ 0.40 &  3 \\
                &               4 & 0.14 $\pm$ 0.25 &  86 \\
                &               5 & 0.22 $\pm$ 0.50  &  2 \\
\hline
   $\kappa$-Aql &               1 &0.27 $\pm$ 0.27 & 2 \\
                &               2 &0.23 $\pm$ 0.24 & 76 \\
                &               3 &0.30 $\pm$ 0.22 & 20 \\
                &               4 &0.29 $\pm$ 0.30 & 1 \\
                &               5 &0.30 $\pm$ 0.30 & 1 \\
\hline
      HD~206267 &               1 & $-$0.12 $\pm$ 0.12 &  6 \\
                &               2 & $-$0.10 $\pm$ 0.11 & 94 \\
\hline
      HD~207198 &               1 & 0.28 $\pm$ 0.33 & 35 \\
                &               2 & 0.28 $\pm$ 0.21 & 61 \\
                &               3 & 0.30 $\pm$ 0.72 & 4 \\
\hline
\end{tabular}
\end{table}

\begin{table}
\centering
\caption{Table of the minimum and maximum metallicities and their respective $z$-test significance for the minimum-difference realisation.}
\label{table:min-met-var}
\begin{tabular}{c|c|c|c|c}
\hline \hline
Target  & Min.      & Max.      & Diff.    & \textit{z-test} \\
                 & [M/H]$_i$ & [M/H]$_i$ & (dex)    & ($\sigma$)          \\
\hline
$\theta^1$ Ori C   & $-$0.92 $\pm$ 0.32 & 0.10 $\pm$ 0.15    & 1.02 & 2.5   \\
HD110432                    & 0.37 $\pm$ 0.19    & 0.41 $\pm$ 0.85    & 0.04 & 0 \\
$\chi$ Oph                  & 0.27 $\pm$ 0.37    & 0.30 $\pm$ 0.07    & 0.10 & 0.4   \\
HD154368                    & 0.14 $\pm$ 0.25    & 0.22 $\pm$ 0.23    & 0.08 & 0.2   \\
$\kappa$ Aql                & 0.23 $\pm$ 0.1     & 0.30 $\pm$ 0.13    & 0.07 & 0.5    \\
HD206267                    & $-$0.12 $\pm$ 0.12 & $-$0.10 $\pm$ 0.11 & 0.02 & 0.1 \\
HD207198                    & 0.28 $\pm$ 0.72    & 0.28 $\pm$ 0.33    & 0    & 0   \\
\hline
\end{tabular}
\end{table}

% --------------------------------------------------------------------
   
   \subsection{Abundance patterns for the whole line of sight} \label{subsec: abundance pattern plots}
   
\begin{figure}[htbp]
    \centering
    \includegraphics[width=0.4\textwidth]{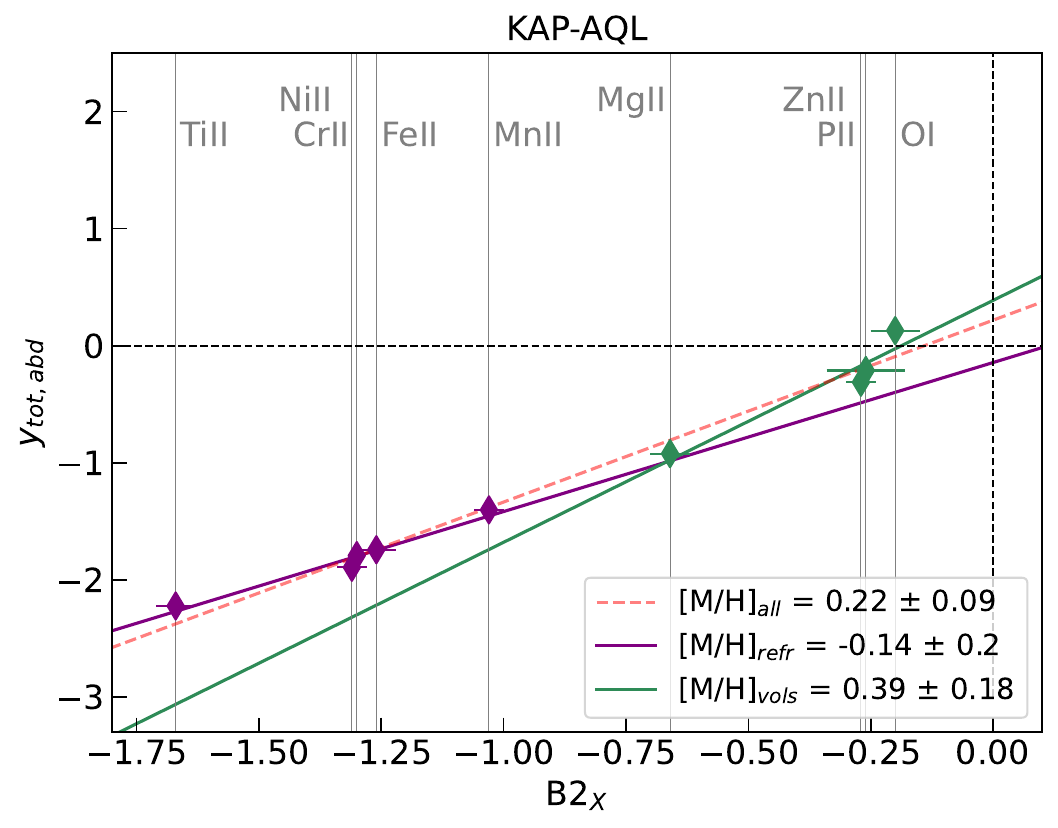}
    \caption{Abundance pattern for the total line of sight towards $\kappa$ Aql. The red dashed line shows the fit to all of the metals, the purple is the fit to the more refractory metals, and the green is fit to the more volatile. }
    \label{fig:kap-aql-abundance-pattern}
\end{figure}

Fig. \ref{fig:kap-aql-abundance-pattern} shows the abundance patterns for full lines of sight (i.e. combining all velocity components) towards our $\kappa$ Aql. The remaining targets are included in Appendix \href{https://zenodo.org/records/14793736}{F}. We perform three different linear fits for most of the targets: (1) fits to all the metals (red dashed line), (2) fits to only the more refractory metals (\ion{Ti}{II}, \ion{Ni}{II}, \ion{Cr}{II}, \ion{Fe}{II}, \ion{Mn}{II}) in purple, and (3) fits to only the more volatile metals (\ion{Mg}{II}, \ion{Ge}{II}, \ion{Zn}{II}, \ion{P}{II}, \ion{O}{I}) in green. We define a limit between more refractory and more volatile elements around a value of $B2_X \sim 0.8$. For $\rho$ Oph A we report the fit to the only two reliable column densities. For $\theta^1$ Ori C we only include results from the fit to all metals and to the more volatile because we find the fit to the more refractory metals yields unreliable results. This could be caused by Fe-group (Ni, Cr, Fe, Mn) over-abundance in this ISM cloud, or by peculiar Ti depletion \citep[which is also seen in lines of sight towards other stars in this region][]{Ritchey+2023}. In general, it is clear that the abundance pattern fits to all the metals are not linear, and that there is an upturn of the more volatile metals. This is likely due to the presence of a mix of gases along the lines of sight with different levels of depletion, and possibly different metallicities.  
 
 The purpose of performing linear fits to only the refractory/volatile elements is to show that we find that these two strategies measure different metallicities for the same lines of sight. For a given abundance pattern, fits to only the more refractory metals systematically produce lower metallicities, while fits to only the more volatile metals result in metallicities that are more similar to fits to all the metals. If low-metallicity gas exists along the line of sight, the refractory metals are the only ones sensitive to it because the volatile elements are always dominated by the omnipresent high-metallicity gas. On the other hand, the refractory elements are mostly highly depleted in the high-metallicity gas, and their abundance can instead be dominated by the low-depletion, low-metallicity gas. It is also possible to have high metallicity gas and low dust content, for example due to dust destruction. However, our approach is agnostic to metallicity because we directly measure the dust depletion of individual components, without assumptions on the metallicity. We may expect the presence of low-metallicity gas in the ISM because there is evidence of accreting low-metallicity gas, for example from HVCs \citep{Fox&Dave2017_gasaccretion}.

\begin{figure}
        \centering
        \includegraphics[width=\hsize]{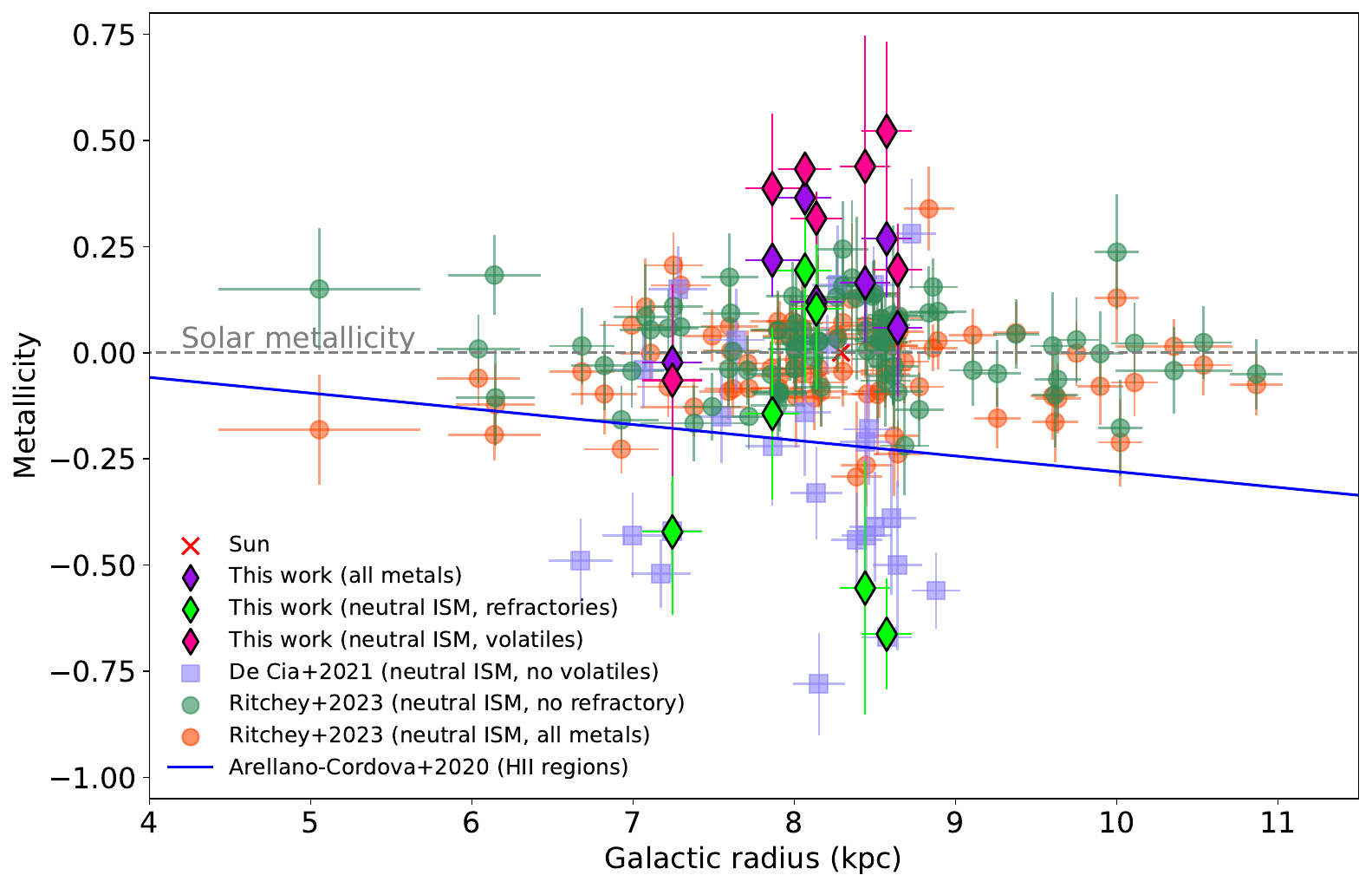}
        \caption{Metallicity calculated over the full lines of sight as a function of distance from the Galactic center plotted for this work compared with literature results. The red diamonds are the results from the fit to abundance patterns for all metals, the green and purple diamonds are fits to the refractory and volatile metals respectively. We do not include the fits for $\rho$ Oph A because of the limited data. The fit to the refractory metals for $\theta^1$ Ori C is 1.7, which we do not consider as physical, so it is excluded in this plot. }
        \label{fig:metallicity-literature}
\end{figure}

 Table \ref{tbl:abundance-patterns-results+literature} summarises these results. In Fig. \ref{fig:metallicity-literature} we compare our results with those from \citet{DeCia+2021, Ritchey+2023}. We include the metallicity gradient based on \ion{H}{II} regions from \citet{Arellano-Cordova+2020_abundances-ionised}.
 
 These results confirm that using abundance patterns to calculate the metallicity of gas along full lines of sight is an approximation, and that component-by-component analyses are important for gaining a holistic measurement of the chemical enrichment in the ISM.

\begin{table*}[]
\begin{tabular}{c|c|c|c|c|c|c|c|c}
\hline \hline
Target & No.   & Groups & [M/H]$_{\mathrm{tot}}$ & [Zn/Fe]$_{\mathrm{fit,tot}}$ & [M/H]$_{\mathrm{ref}}$ & [M/H]$_{\mathrm{vol}}$ & [M/H]$_{\mathrm{DC21}}$ & [M/H]$_{\mathrm{R23}}$ \\
                 & comps & of comps   &                                 &                                       &                                 &                                 &                            &   all metals                              \\ \hline 
 $\theta^1$ Ori C         & 16             & 4                   &  0.06 $\pm$ 0.20                 & 1.46 $\pm$ 0.20                       & --                 & 0.20 $\pm$ 0.10                 & $-$0.50 $\pm$ 0.20            & --              \\
 HD 110432             & 6              & 2                   & 0.36 $\pm$ 0.07                 & 1.78 $\pm$ 0.06                       & 0.19 $\pm$ 0.27                 & [0.43]                          & $-$0.14 $\pm$ 0.15           &                                 \\
 $\rho$ Oph A          & 15             & 2                   & [0.36]                          & [1.91]                                & --                              & --                              & $-$0.78 $\pm$ 0.12         &   $-$0.036 $\pm$ 0.094 \\
 $\chi$ Oph            & 19             & 4                   & 0.12 $\pm$ 0.06                 & 1.66 $\pm$ 0.06                       & 0.1 $\pm$ 0.19                  & 0.36 $\pm$ 0.07                 & $-0.33$ $\pm$ 0.11            &      --         \\
 HD 154368            & 11             & 5                   & $-$0.02 $\pm$ 0.08                & 1.45 $\pm$ 0.08                       & $-$0.42 $\pm$ 0.2               & $-$0.06 $\pm$ 0.23              & $-$0.42 $\pm$ 0.12         &                                 \\
 $\kappa$ Aql          & 16             & 5                   & 0.22 $\pm$ 0.09                 & 1.60 $\pm$ 0.07                       & $-$0.14 $\pm$ 0.2               & 0.37 $\pm$ 0.18                 & $-$0.22 $\pm$ 0.14            &                                 \\
 HD 206267             & 7              & 2                   & 0.16 $\pm$ 0.14                 & 1.51 $\pm$ 0.12                       & $-$0.55 $\pm$ 0.3               & 0.44 $\pm$ 0.31                 & $-0.21$ $\pm$ 0.12            & $-$0.097 $\pm$ 0.067 \\
 HD 207198             & 6              & 2                   & 0.27 $\pm$ 0.14                 & 1.56 $\pm$ 0.12                       & $-$0.66 $\pm$ 0.13              & 0.52 $\pm$ 0.21                 & $-0.67$ $\pm$ 0.08            & $-$0.05 $\pm$ 0.067
        \\ \hline
\end{tabular}      
\caption{Results from the abundance patterns for each target. We include the metallicities calculated over the full lines of sight obtained from only fitting to the refractory metals and only fitting to the volatile metals, as well as a comparison to the literature. [M/H]$_{\mathrm{DC21}}$ and [M/H]$_{\mathrm{R23}}$ are metallicities from \cite{DeCia+2021} and \cite{Ritchey+2023} respectively. The values in square brackets are from fits done to only two data points. }
\label{tbl:abundance-patterns-results+literature}
\end{table*}

%======================================================================
%======================================================================
%======================================================================

\section{Conclusions} \label{sec:conclusions}
   
   In this work we study individual (groups of) components along eight lines of sight within 1.1 kpc of the Sun. We use metal patterns and the relative method to determine the amount of dust depletion [Zn/Fe]$_{\mathrm{fit, }i}$ in each component. We investigate possible metallicities for these individual components by simulating 99, (4~851, 156~849, 1~000~000) distributions of the total hydrogen gas between 2 (3, 4, 5) components. Our results are summarised below.
   \begin{enumerate}
      
      \item We find components with very high levels of dust depletion (up to [Zn/Fe]$_{\mathrm{fit}} = 2.03 \pm 0.03$ dex or $F^* = 1.30$) compared to the full line-of-sight analysis conducted by \cite{DeCia+2021}, who report a maximum depletion of 1.32 or $F^* = 0.53$. Our study shows that full line-of-sight analyses can wash out the range of dust depletion. This means that it is possible to have much higher levels of depletion in individual clouds compared to the full line of sight. 
      
      \item We find large ranges in the level of dust depletion among components along lines of sight, calculated as the difference between the minimum and maximum depletion, up to a factor of 15 (1.19 dex). 
      
      \item We present a novel method of constraining the metallicities of individual gas components. We use information gained from the metals patterns, i.e. that the $y$-intercept is the sum of the hydrogen gas fraction and the metallicity $a_i$ = $\log {N(\mathrm{H})_i +}$ [M/H]$_i$, with simulated $\log{N(\mathrm{H})_i}$ distributions to calculate [M/H]$_i$ for individual components. We implement an upper metallicity limit of 0.5 dex based on results from \cite{Nepal+2024_MW-metallicity}. From our simulations and under this assumption, we are able to constrain metallicities for some components to be within an interval of $\sim$0.3 dex. 
      
      \item Our simulations indicate that, in five out of eight lines of sight, the component with the highest level of dust depletion contains the majority of the hydrogen gas, and likely has super-Solar metallicity.

      \item For the line of sight towards $\theta^1$ Ori C, we find a group of components (group 2 out of 4 groups) with high dust depletion [Zn/Fe]$_{\mathrm{fit, }2} = 1.52 \pm 0.25$, the bulk of the hydrogen gas 37\% $<f_2 <$ 97\%, and super-Solar metallicity [M/H]$_2 \geq$ 0.08 dex. This group of components could be attributed to the nebula itself \citep{Price+2001}. Further, we find that group 4 has a low level of depletion [Zn/Fe]$_{\mathrm{fit, }4} = 0.60 \pm 0.21$, and likely low metallicity, at least for the case with the minimum difference in metallicity.

      \item From our realisations, we find that the minimum variation in metallicity is $< 0.15$ dex for all targets, with the exception of $\theta^1$ Ori C, which we find to be 1.02 dex. Although the metallicity of individual gas components components cannot be strictly measured, this methodology allows us to study the minimum variations of metallicity, from component to component.
  
   \end{enumerate}

With the methodology we develop in this paper to constrain the metallicities of individual gas clouds in the Solar neighbourhood, we are able to study the ISM in more detail, and we gain a better understanding of the level of chemical variation in the ISM.

\section{Data Availability}

Appendices D - F are available on Zenodo at this \href{https://zenodo.org/records/14793736}{link}.   

\begin{acknowledgements}
   Edward B. Jenkins passed away just before the submission of this paper. Ed was one of the great fathers of UV astronomy and his scientific work on the ISM was immensely impactful and inspiring. We are grateful to have had the opportunity to work with and learn from him. His legacy will live on. Daniel E. Welty calculated the column densities for the targets $\theta^1$ Ori C, $\chi$ Oph and $\kappa$ Aql. The authors acknowledge and thank him for his significant and valuable contributions to this paper. T.R.-H., A.D.C., J.-K.K., C.K. and A.V. acknowledge support by the Swiss National Science Foundation under grant 185692. This work is based on observations with the NASA/ESA Hubble Space Telescope obtained from Mikulski Archive for Space Telescopes at the Space Telescope Science Institute, which is operated by the Association of Universities for Research in Astronomy, Incorporated, under NASA contract NAS5-26555. We also make use of optical data from the European Southern Observatory, Kitt Peak National Observatory, McDonald Observatory and the Anglo-Australian Telescope. Support for program number 16750 was provided through a grant from the STScI to Princeton University and STScI under NASA contract NAS5-26555. This research has made use of NASA’s Astrophysics Data System.
\end{acknowledgements}

\bibliographystyle{aa}
\bibliography{AandA_MilkyWay}`

\newpage

\appendix

% ===========================================================================

\section{Details of observations}
Table \ref{tbl:observation-details} shows the details of the HST/STIS cycle 29 observations (ID: 16750) used in this work.

\begin{table*}[h] 
\centering
\begin{tabular}{c|c|c|c|c|c|c|c|c}

\hline \hline
Target & Orbits & Aperture   & Exposure time & Max. & Orbits & Aperture   & Exposure time & Max. \\
 & E140H/1271 &   & (s) & SNR & E230H/2113 &   & (s) & SNR \\ \hline
$\theta^1$ Ori C  & 1                          & 31X0.05NDB & 1786          &  48           & 1                          & 31X0.05NDC & 2508          & 65          \\
HD~110432       & 1                          & 0.1X0.03   & 2828          & 74          & 1                          & 0.1X0.03   & 2040          & 52          \\
$\chi$ Oph      & 1                          & 31X0.05NDB & 2536          & 49          & 1                          & 31X0.05NDA & 1866          & 49          \\
$\rho$ Oph A    & --                         &    --        &    --           &   --          & 1                          & 0.2X0.06   & 1786          & 59          \\
HD~154368       & 2                          & 0.2X0.2    & 5395          & 21          & 2                          & 0.2X0.2    & 5143          & 20          \\
$\kappa$ Aql    & 1                          & 31X0.05NDB & 2508          & 61          & 1                          & 31X0.05NDA & 1786          & 53          \\
HD~206267       & --                         &   --         &    --           &   --          & 1                          & 6X0.2      & 2018          & 33          \\
HD~207198       & --                         &   --         &    --           &   --          & 1                          & 6X0.2      & 2030          & 41          \\ \hline
\end{tabular}
\caption{Table of the observing details for HST cycle 29 program ID: 16750. The reported maximum SNR is per pixel. Additional archival data are used in this work as described in Section \ref{sec:data}.}
\label{tbl:observation-details}
\end{table*}

% ===========================================================================
\section{Methodology test} \label{appendix:method-test}
We test our methodology described in Section \ref{sec:methods} by adding an extra component with known metallicity and total hydrogen column density to the line of sight towards HD~206267. This component, chosen from the SMC, helps us verify that our method can accurately recover the metallicity difference between the Milky Way and the SMC. We use data from \cite{DeCia+2024} for AzV~78, who report [Zn/Fe]$_{\mathrm{fit}} = 0.62 \pm 0.12$, [M/H] = $-0.82 \pm 0.17$ and $\log{N(H)} = 21.7 \pm 0.06$. We choose the sight line towards HD~206267 because it has only two components with relatively low dust depletion, reducing the degeneracy between the components. This makes the hydrogen gas fraction for the SMC component $f_{\mathrm{SMC}} = 74\%$.

\begin{figure}
    \centering
    \includegraphics[width=0.5\textwidth]{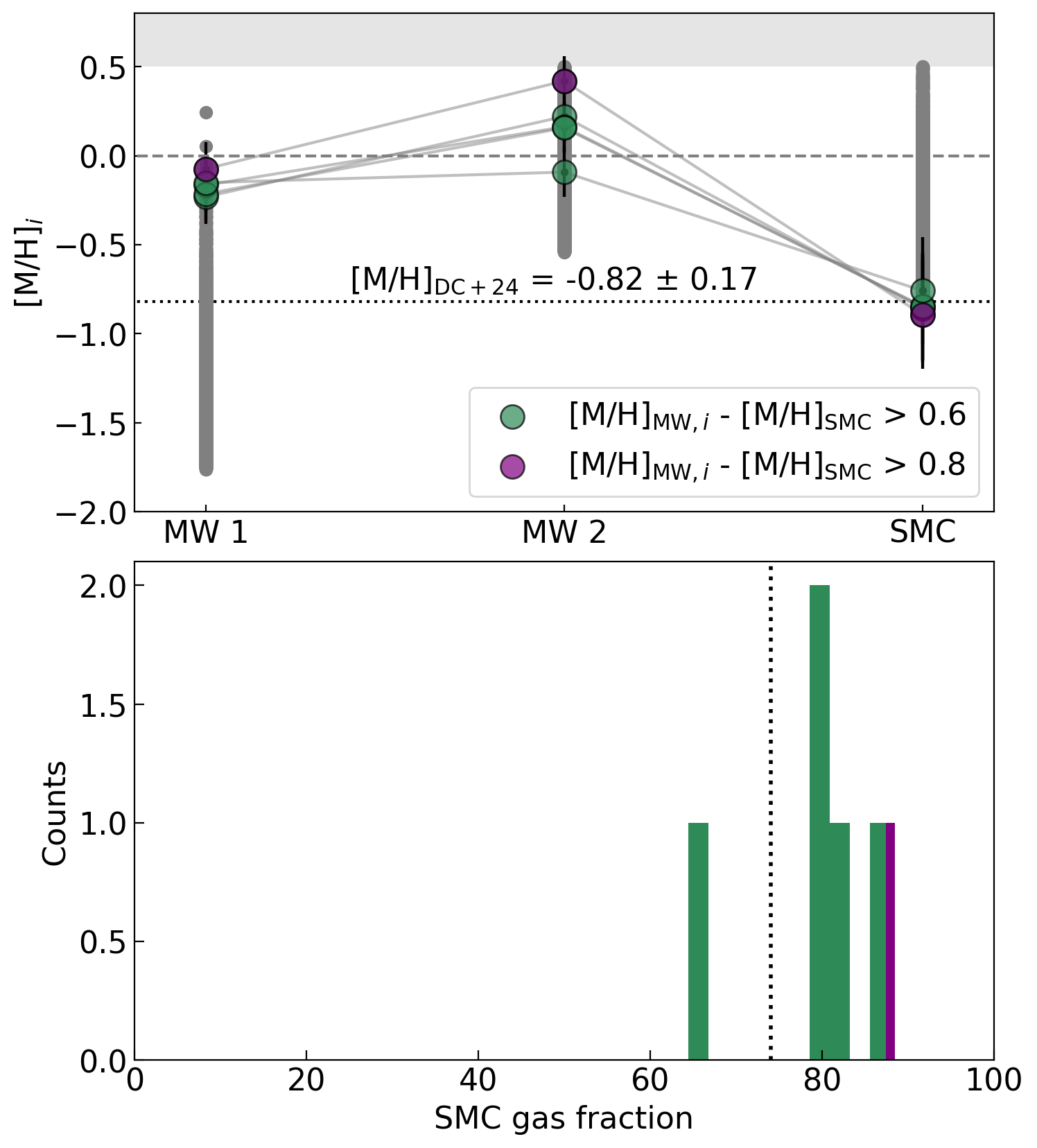}
    \caption{Results from testing of our method in Section \ref{sec:simulating clouds} showing we reproduce the large range in metallicity between the Milky Way and the SMC for 5 out of the 4865 realisations. In the top panel we plot all of the `most-likely' realisations in grey and over-plot those which recover a 0.6 dex difference in metallicity between both of the Milky Way clouds and the SMC cloud in green. The histogram in the bottom panel show the total hydrogen gas distributions for these realisations.}
    \label{fig:test-model-HI}
\end{figure}

For 5 out of the 4865 most likely realisations (i.e. metallicities of all components are $< 0.5$ dex) from this exercise, we recover a difference in metallicities of $> 0.6$ dex between both Milky Way clouds and the SMC (i.e. [M/H]$_{\mathrm{MW 1}} - $ [M/H]$_{\mathrm{SMC}} > 0.6$ dex and [M/H]$_{\mathrm{MW 2}} - $ [M/H]$_{\mathrm{SMC}} > 0.6$ dex). These realisations also reproduce the SMC metallicity of $-0.82 \pm 0.17$. There is one realisation that produces a difference of 0.8 dex, and none produce larger than than this. In the former case the SMC metallicity is also reproduced. Fig. \ref{fig:test-model-HI} shows the 5 realisations with metallicity differences $>$ 0.6 dex in the green points and the 1 realisation that produces a difference of $> 0.8$ dex in purple, plotted over all of the `most likely' realisations in grey (see Section \ref{subsec: variations in metallicity}). The metallicity of the SMC cloud is shown by the horizontal dotted black line. The gas fraction histogram in the bottom panel shows that this exercise somewhat reproduces the SMC gas fraction. 

% ===========================================================================

\end{document}